\documentclass[a4paper,11pt]{article}
\usepackage{geometry}
\usepackage[numbers]{natbib}
\usepackage[T1]{fontenc}
\usepackage{amssymb,amsthm,bm,mathrsfs,mathtools}
\usepackage[dvipsnames]{xcolor}
\usepackage[bookmarks=false,colorlinks=true,allcolors=blue]{hyperref}
\usepackage{graphicx,tabularx}

\newcommand{\ii}{\mkern1mu \mathrm{i}\mkern1mu} 
\newcommand{\ee}{\mathrm{e}} 
\newcommand{\tp}{{\mkern-1mu \mathsf{T}}} 
\renewcommand{\Re}{\operatorname{Re}}
\DeclareMathOperator{\diag}{diag}
\DeclareMathOperator{\tr}{tr}
\DeclareMathOperator{\sig}{sgn}
\DeclarePairedDelimiter\abs{\lvert}{\rvert}
\DeclarePairedDelimiter\norm{\lVert}{\rVert}
\DeclarePairedDelimiter\avg{\langle}{\rangle}
\DeclarePairedDelimiter\ket{\vert}{\rangle}
\DeclarePairedDelimiter\bra{\langle}{\vert}
\DeclarePairedDelimiterX{\braket}[2]{\langle}{\rangle}{#1\delimsize\vert\mathopen{} #2}
\DeclarePairedDelimiterX{\matele}[3]{\langle}{\rangle}{#1\delimsize\vert\,\mathopen{}#2\,\delimsize\vert\mathopen{}#3}
\newcommand{\dyad}[1]{\ket{#1}\!\bra{#1}}
\newcommand{\repr}{\mathrel{\widehat{=}}}

\DeclarePairedDelimiter\bond{\langle}{\rangle}

\newcommand{\Eqref}[1]{Eq.~\!\eqref{#1}}
\newcommand{\Eqsref}[1]{Eqs.~\!\eqref{#1}}
\newcommand{\Figref}[1]{Fig.~\!\ref{#1}}
\newcommand{\mr}[1]{\mathrm{#1}}
\newcommand{\mb}[1]{\mathbf{#1}}

\newcommand{\wt}[1]{\widetilde{#1}}
\newcommand{\bv}[1]{\bm{\vec{#1}}}
\newcommand{\CZ}{\ensuremath{\mathrm{CZ}}}

\newcommand{\Fid}{\mathrm{F}}
\newcommand{\InF}{\mathrm {InF}}
\newcommand{\Uqb}{{U}_\mr{q}}
\newcommand{\Uideal}{{U}_\mr{q}^{(1)}}
\newcommand{\Ez}{\varepsilon}

\usepackage{authblk}
\title{Multi-qubit DC gates over an inhomogeneous array of quantum dots}
\author[1]{Jiaan Qi \thanks{Corresponding:\ qija@baqis.ac.cn}}
\author[1]{Zhi-Hai Liu}
\author[1,2]{Hongqi Xu \thanks{Corresponding:\ hqxu@pku.edu.cn}}
\affil[1]{Beijing Academy of Quantum Information Sciences, Beijing 100193, China}
\affil[2]{Beijing Key Laboratory of Quantum Devices and School of Electronics, Peking University, Beijing 100871, China}
\date{}

\begin{document}
\maketitle

\begin{abstract}
The prospect of large-scale quantum computation with an integrated chip of spin qubits is imminent as technology improves. This invites us to think beyond the traditional 2-qubit-gate framework and consider a naturally supported ``instruction set'' of  multi-qubit gates. In this work, we systematically study such a family of multi-qubit gates implementable over an array of quantum dots by DC evolution. A useful representation of the computational Hamiltonian is proposed for a dot-array with strong spin-orbit coupling effects, distinctive $g$-factor tensors and varying interdot couplings. Adopting a perturbative treatment, we model a multi-qubit DC gate by the first-order dynamics in the qubit frame and develop a detailed formalism for decomposing the resulting gate, estimating and optimizing the coherent gate errors  with appropriate local phase shifts for arbitrary array connectivity. Examples of such multi-qubit gates and their applications in quantum error correction and quantum algorithms are also explored, demonstrating their potential advantage in accelerating complex tasks and reducing overall errors.  
\end{abstract}

\maketitle 

\clearpage
\section{Introduction}
Semiconductor quantum dots are promising physical platforms for universal quantum computing  \cite{Chatterjee2021Semiconductor,Vaughan2023platform}. In this widely conceived scheme, the spins of electrons (or holes) are confined with artificial structures of nanoscale, and are selectively manipulated and brought into interactions using accurate electromagnetic signals \cite{Burkard2023Semiconductor,Fang2023Recent}. 
Owing to their miniature size and  compatibility with modern semiconductor fabrication techniques  \cite{Zwerver2022Qubits,Koch2025Industrial}, semiconductor spin qubits have great potential for coherently incorporating a large quantity of qubits in a single chip, a crucial requirement for solving useful quantum computational tasks.  
Over recent years, significant advancements in key performance indicators such as  coherence lifetime, single- and 2-qubit gate fidelity have been made \cite{Stano2022Review,Mills2022HighFidelity,Noiri2022Fast,Xue2022Quantum}.
In terms of scaling-up the system size, a viable way is to employ an extensible array of quantum dots, as recently demonstrated for two-dimensional  crossbar arrays of Germanium hole qubits \cite{Hendrickx2021fourqubit,Lawrie2023Simultaneous,Borsoi2024Shared,zhang2023universal}. 
This configuration is also compatible with the surface code, the golden framework for creating fault-tolerant quantum computers  \cite{fowler2012surface,Acharya2025Quantum}.

\begin{figure*}[htbp]
 \centering
 \includegraphics*[width=0.9\linewidth]{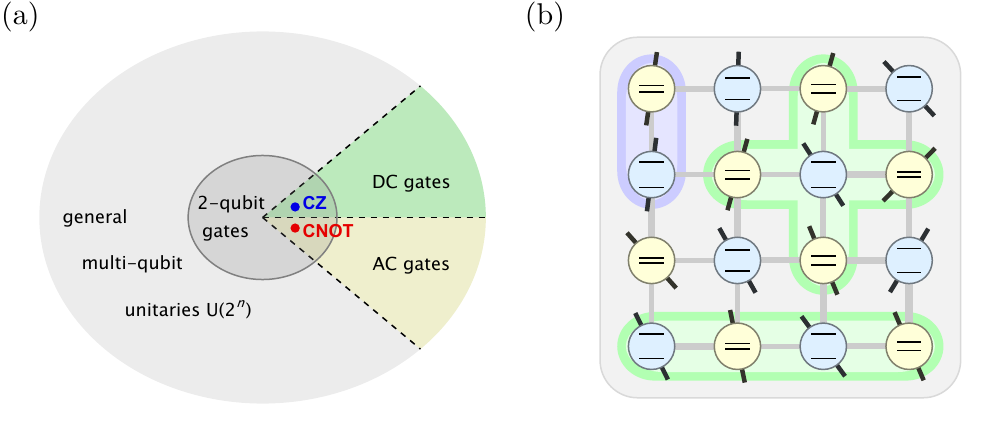}
 \caption{(a) A classification diagram of quantum gates.
While a general multi-qubit unitary in the outer ring can be composed from universal 2-qubit gates in the inner disk,  only a small number of these gates are directly implementable. 
The set of naturally accessible gates is determined by the interaction Hamiltonian and forms a cone (the shaded fan) that extends beyond the 2-qubit limit into  multi-qubit domain. 
For spin qubits in particular, we can distinguish the DC class and the AC class, with the familiar CZ and CNOT gate as the respective 2-qubit member. 
This work generalizes the CZ/CPhase gate to a broader set of primitive multi-qubit DC gates for spin qubits. (b) A schematic plot of the modeled spin-qubit quantum chip, which is composed of an inhomogeneous array of quantum dots with varying Zeeman splitting energies, quantization axes and interdot coupling strengths. Apart from the 2-qubit CPhase gates (in the blue rectangle), it is revealed that some multi-qubits gates (in green regions) can also be naturally implemented  with high fidelity on such array. }
 \label{fig:class}
\end{figure*}

Current existing studies on entangling gates have primarily focused on 
universal two-qubit gate such as the controlled-not (CNOT) gate for implementing two-qubit logic \cite{Wu2023Hamiltonian}. Nevertheless, any quantum circuit may be carried out with an alternative set of universal gates most convenient for the physical platform. For example, the CNOT gate are often better implemented by combing the controlled-phase (CPhase)/controlled-Z (CZ) gate with other single-qubit gates for spin qubits \cite{Xue2022Quantum}, and through a combination of the cross-resonance gate with single-qubit gates for superconducting qubits \cite{Chow2011Simple}. More generally, one may classify all quantum gates according to their accessibility for the hardware, as illustrated in \Figref{fig:class}(a). The set of primitive gates, namely the quantum gates achievable within a single control segment, constitutes a cone in the space of all unitary transformations \cite{Bengtsson2009Geometry}. For spin qubits, these primitive quantum gates can be divided into the DC gates and the AC gates according to the natural of the control signals, with the CZ gate and CNOT gate being the respective 2-qubit members \cite{Russ2018Highfidelity,Watson2018programmable,Zajac2018Resonantly}. 
One step further, for multi-qubit systems we should also drop the two-qubit restriction and consider the \emph{multi}-qubit gates supported by the {hardware}. These multi-qubit gates are directly derived from the many-body interaction Hamiltonian and can be efficiently realized at high accuracy. In analogy to the instruction set of a classical CPU, such architecture-dependent   multi-qubit gates constitute an instruction set for the QPU (quantum processing unit).
Quantum computing circuits can and should be aptly composed from these primitive gates for boosting efficiency, simplifying control, and reducing overall logical errors \cite{Gu2021Fast}.

Owing to their apparent benefits, research interests on multi-qubit gates have recently emerged across various architectures.  Notable examples include those in Rydberg atoms \cite{Isenhower2011Multibit,Khazali2020Fast,Yu2020Spheroidalstructurebased,Young2021Asymmetric}, superconducting circuits \cite{Gu2021Fast,Chancellor2017Circuit,Liu2018Onestep,Bahnsen2022Application} and trapped-ions \cite{Rasmussen2020Singlestep,Espinoza2021highfidelity}. 
For spin-based systems, the three-qubit Toffoli gate has been recently proposed and verified by applying resonant microwave pulses \cite{Gullans2019Protocola,Takeda2022Quantum}. Three-qubit DC gates also also been explored for a linear array from the perspective of superexchange couplings \cite{Rancic2017Ultracoherent,Gullans2019Protocola}.
In particular, a novel 3-qubit gates induced by chiral interaction on a triangle have also been theoretically explored in a recent paper \cite{Nguyen2025Singlestep}. 
Notably, a series of studies on linear quantum dot chains have revealed coherent ``superexchange'' oscillations of the edge states that exhibit long-range coupling across multiple dots \cite{Qiao2020Coherent,Qiao2021LongDistance,Knorzer2022Longrange}. These discoveries clearly hint the existence of a rich number of multi-qubit gates for spin-qubit chips. Despite these existing efforts in addressing multi-qubit gates in specific configurations, a bird-eye view on the overall landscape of multi-qubit gates for spin qubits is still lacking and much needed given the trend of ever-expanding qubit counts in a single chip.

In this paper, we attempt to partly fill this important research gap by extending the well-established  2-qubit CZ/CPhase gate to a broader family of multi-qubit DC gates naturally applicable to quantum-dot arrays. 
Capturing their structural similarity,  we are able to describe the general structures and properties of multi-qubit DC gates applicable to quantum dots.
 At this level of research scope, we aim at addressing the following questions,
\begin{enumerate}
  \item What are the possible multi-qubit gates that are applicable?
  \item What are the expected error rates of these multi-qubit gates? In particular, how do they compare to the equivalent two-qubit gates?
  \item What are the practical advantages of the proposed multi-qubit gates?
\end{enumerate}
These questions are of key importance in developing powerful instruction sets for quantum processors.  To address these questions,  an analytical framework based on hierarchical perturbation is developed for analyzing the gate dynamics and gate fidelities. Our work unravel a large segment of previously uncharted territory in the gate space, as indicated in the green-shaded fan in \Figref{fig:class}(a). Hopefully,  these preliminary powerful can pave way for sophisticated quantum gate designs in large-scale spin-qubit chips.

The spin-qubit chip we considered in this study constitutes an array of quantum dots arranged in 1 or 2 spatial dimensions.  
The system is kept at low temperature and works in the half-filling regime, i.e., only one carrier (electron or hole) per dot.  An external static magnetic field lifts the orbital degeneracies and defines a spin qubit on each dot. Notably, by assuming an inhomogeneous array of quantum dots, our theory addresses some practically relevant aspects often neglected in earlier models \cite{DiVincenzo2000Universala,Acuna2024Coherenta}. First, different quantum dots can possess distinctive Land\'e $g$-factors (typically tensors), which could lead to spin precession when carriers tunnel across dots \cite{Zwerver2023Shuttlinga,vanRiggelen-Doelman2024Coherent}.  Next, the device can possess significant spin-orbit coupling.  Spin-orbit coupling is necessary to allow fast manipulation of spin states with electric signals \cite{Bosco2021Hole,Froning2021Ultrafast,Kawakami2016Gate}, however it inevitably results in anisotropic exchange coupling between neighboring qubits as opposed to the usual Heisenberg exchange coupling \cite{Li2014Anisotropic,Qi2024Spin}.
Furthermore, due to inhomogeneity in local potential,  the coupling strength can differ across bonds (i.e., in different pairs of exchange-coupled quantum dots). We illustrate such a model device and some of the applicable multi-qubit gates in \Figref{fig:class}(b).

This paper is organized as follows. 
In section~\ref{sec:comH}, we describe the physical setup of our model device and derive the effective Hamiltonian for the computational manifold. In section~\ref{sec:intriG}, we look into a set of quantum gates that are intrinsically attainable on quantum dot arrays. A theorem is established for decomposing multi-qubit gates with global and local phase gauge depending on the connectivity. We estimate the coherent fidelity of these multi-qubit gates and prove that they can be reliably implemented following a certain device design.
In section~\ref{sec:examples}, we explicitly consider some examples of DC multi-qubit gates and discuss their potential applications in quantum computational tasks.  
We summarize and discuss further research directions in section~\ref{sec:conclusion}.

\section{The computational Hamiltonian}\label{sec:comH}

The Fermi-Hubbard model is a good starting point for describing a half-filling quantum dot array \cite{Yang2011Generic}. 
The ground-level bound states of all the dots can be normalized to form a low-energy basis $\{\ket{\Phi_{j\sigma}}\}$, where $j$ is the dot index and $\sigma\in\{\uparrow,\downarrow\}$ indicates the spin.
Introducing the annihilation $a_{j\sigma}$, creation $a^+_{j\sigma}$ and number $n_{j\sigma}=a^+_{j\sigma}a_{j\sigma}$ operators for these basis states, one can write down the second-quantized Hamiltonian,
\begin{subequations}\label{eq:FH-Hamiltonian}
\begin{align}
&H_\mr{FH} = H_\mr{dot} + H_\mr{tun}, 
\\[1ex]
 & H_\mr{dot} = \sum_{j}\sum_{\sigma} \Bigl[\bigl(\mu_j + {\sig(\sigma_j)}\,\frac{1}{2}\Ez_{j}\bigr) n_{j\sigma} + \frac{1}{2} U_j  n_{j\sigma} n_{j\bar\sigma}\Bigr], \\
 &H_\mr{tun} =
\sum_{\bond{j,k}}\sum_\sigma \left( t^{jk}_{\sigma\sigma} a^{+}_{j\sigma} a_{k\sigma} 
+ t^{jk}_{\sigma\bar\sigma} a^{+}_{j\sigma} a_{k\bar\sigma}\right),
\label{eq:Htun}
\end{align}
\end{subequations}
where $\bar\sigma$ stands for the opposite orientation of $\sigma$,
with the spin sign defined by $\sig(\uparrow\downarrow)=\pm 1$.
$H_\mr{dot}$ describes the energy cost for filling charge carriers onto the dots, with the local potential $\mu_j$, Zeeman splitting energy $\Ez_j$ and the charging energy  $U_j$.   $H_\mr{tun}$ describes the interdot tunneling of the charge carriers. 
The first summation in \Eqref{eq:Htun}  is performed over all pairs of adjacent dots $\bond{j,k}$ coupled via exchange tunneling (we refer such a pair as ``bond'' in later texts).

The tunneling Hamiltonian [\Eqref{eq:Htun}] can be split into four ``spin-conserved'' tunneling terms  ($a^{+}_{j\sigma} a_{k\sigma}$) and four ``spin-flipped'' tunneling terms  ($a^{+}_{j\sigma} a_{k\bar\sigma}$) for each bond. Intuitively, the spin-flipping process is often attributed to an effective spin-orbital field when charge carriers tunnel across dots \cite{zhang2023universal,Geyer2024Anisotropic}. 
Microscopic models have been used to derive the expressions  for the preceding tunneling coefficients \cite{Froning2021Strong,Liu2018Control}. 
Meanwhile, other factors such as differences in the dot quantization axes and many-body dipole interactions can all contribute to spin precession \cite{Zwerver2023Shuttlinga}. 
Despite these possible complications, a set of relations among the  tunneling coefficients can be still obtained based solely on a  symmetry relating to time-reversal. Under an external magnetic field $\bm{B}$, the time-reversal symmetry of the system is strongly broken due to the Zeeman term in $H_\mr{dot}$. However, provided that the interdot tunneling coefficients changes insignificantly with $\bm{B}$, the tunneling Hamiltonian $H_\mr{tun}$ is still time-reversal symmetric. Through such ``weak'' time-reversal symmetry, combined with Hermicity of the Hamiltonian, one obtains 
\begin{equation}\label{eq:tun-coeffs}
\begin{aligned}
t^{jk}_{\uparrow\uparrow} &=\phantom{-} (t^{jk}_{\downarrow\downarrow})^{*} 
= \phantom{-} t^{kj}_{\downarrow\downarrow} =( t^{kj}_{\uparrow\uparrow})^* &&\equiv \,
t_{jk},\\ 
t^{jk}_{\uparrow\downarrow} &= -(t^{jk}_{\downarrow\uparrow})^{*} 
= -t^{kj}_{\uparrow\downarrow} =( t^{kj}_{\downarrow\uparrow})^* &&\equiv \,
s_{jk}.
\end{aligned}
\end{equation}
Here we introduce for each bond the spin-conserved ($t_{jk}$) and the spin-flipped ($s_{jk}$) tunneling coefficients, which are complex numbers in general.  We note that with the time-reversal symmetric $H_{\mr{tun}}$, a 4-dimensional real spin-orbit vector can be defined~\cite{Danon2009Pauli}, echoing the 2 complex coefficients here.
The weak time-reversal symmetry holds provided that the Zeeman splitting energy is much smaller than the interdot potential barrier, which is a valid approximation for a typical spin-qubit chip where the on-site spin splitting energy  ($\sim1~\mu$eV) is much smaller than the orbital energy ($\sim 1{-}10$ meV) \cite{Li2013Controlling}. 
A detailed derivation and analysis of this result is provided in Appendix~\ref{app:tunnel}. 

Quantum information is encoded by the low-energy half-filling states of the array. For an array of $N$ dots, we denote such a state by 
\begin{equation}\label{eq:comp-state}
\begin{aligned}
\ket{n}\equiv\ket{\sigma_{n_1},\sigma_{n_2},\cdots,\sigma_{n_{N}}}, \quad \sigma_{n_j}\in\{\uparrow,\downarrow\},
\end{aligned}
\end{equation}
where the binary component $\sigma_{n_j}\in\{\uparrow,\downarrow\}$ represents the spin state at dot $j$. These states constitute a basis for the $2^N$-dimensional manifold where quantum computation takes place and are also referred to as the computational states.   It should be stressed that \Eqref{eq:comp-state} is a  Fock space shorthand for the underlying many-body wavefunction which is totally antisymmetric with respect to single-body basis states. 

The immediate state after tunneling involves two spins occupying the same dot that is not energetically favored but allowed briefly by quantum mechanics. This virtual tunneling process gives rise to direct exchange interaction, the dominant interqubit coupling mechanism assumed for our device.  Applying the Schrieffer-Wolf transformation to $H_\mr{FH}
$ followed by an  projection onto the computational manifold, we obtain the effective Hamiltonian
\begin{equation}\label{eq:Heff}
  H = H_0 + H_\mr{ex}= \sum_j \frac{1}{2} \Ez_j \sigma_j^{\mr{Z}} \,-\sum_{\mathclap{w=\langle{j,k}\rangle}} J_w \ket{\xi_w}\bra{\xi_w},
\end{equation}
where the first part $H_\mathrm{0}$ defines the energy splitting of the qubits, which is the summation of the qubit energy splittings $\Ez_j$ 
along the Pauli operator $\sigma_j^\mr{Z}= {\ket{\uparrow}\!\bra{\uparrow}_j}-{\ket{\downarrow}\!\bra{\downarrow}_j}$
for all the dots. The exchange term $H_\mathrm{ex}$ describes the exchange couplings among the bonds. It is specified by associating with each bond a scalar exchange energy,
\begin{equation}\label{eq:J}
 J_{ jk } = \frac{\abs{t_{jk}}^2+ \abs{s_{jk}}^2}{2}\left(\frac{1}{U_j+\mu_j-\mu_k}+\frac{1}{U_k+\mu_k-\mu_j}\right),
\end{equation}
in addition to an entangled state 
\begin{equation}\label{eq:ent-state}
  \ket{\xi_{w}} = \frac{1}{\sqrt{2}} \Bigl( \tilde{s}_w \ket{\uparrow \uparrow}_{w}
- \tilde{t}_w \ket{\uparrow\downarrow}_{w} + \tilde{t}^*_w\ket{\downarrow\uparrow}_{w}+ \tilde{s}^*_w \ket{\downarrow\downarrow}_{w} \Bigr),
\end{equation}
where $(\tilde t_{w},\tilde s_{w})\equiv (t_{w},s_{w})/\sqrt{\abs{t_{w}}^2+\abs{s_{w}}^2}$ is the pair of normalized dimensionless tunneling coefficients associated with each bond $w$.
Explicit derivations of the computational Hamiltonian is demonstrated in Appendix~\ref{app:compH}.

Here we make several remarks regarding the exchange Hamiltonian in \Eqref{eq:Heff}.  
It is nothing new but an alternative, entanglement state based representation of the generalized Heisenberg Hamiltonian.
If all spin-flipping tunneling coefficients are set to be zero, $\{\ket{\xi_w}\}$ will become singlet states of the associated qubit pairs and we recover the familiar Heisenberg exchange interaction with  isotropic coupling $H_\mr{ex} = \sum_{jk} J_{jk} \, \bm{S}_j\cdot\bm{S}_k$.
With the additional spin-flipping channels, which can arise from both the spin-orbital coupling effects and differences in local quantization axes, the exchange coupling becomes \emph{anisotropic} and can be represented as $H_\mr{ex}= \sum_{jk} \bm{S}_j \mathcal{J}_{jk} \bm{S}_k$, where  $\mathcal{J}_{jk}$ is generally a tensor that preserves axial symmetry. See Appendix \ref{app:axisym} for an explicit proof.
Such axially-symmetric form are widely conceived for describing anisotropic exchange interaction \cite{Li2014Anisotropic,Hetenyi2020Exchange}.
Representing anisotropic exchange coupling with entangled states is physically intuitive, mathematically compact and allows easier treatment of multi-qubit gates that will be developed in the following sections. 
On the other hand, we should note that in deriving the effective Hamiltonian only direct tunnelings between nearest-neighbors are taken into account. It is possible, and an interesting research topic, to also include higher-order tunneling effects that lead to effective three-body or four-body interactions \cite{Hsieh2012Herzberg,Milivojevic2017Effectivea}. Notably, a triple-dot system with strong chiral interaction is considered for implementing a 3-qubit gate in a recent study~\cite{Nguyen2025Singlestep}.
 In the working regime of our modeled spin-qubit device, the charging energy is assumed to be  much greater than the tunneling energy, $U\gg \abs{t}$. As a result, two-body interactions, on the order of $\abs{t}^2/U$, is the leading order mechanism for inter-qubit coupling. In comparison, three-body interactions are on the order of $\abs{t}^3/U^2$.  Consequentially these higher-order effects are neglected as small coherent errors in gate implementations.

\section{The multi-qubit gates}\label{sec:intriG}
For systems of two quantum dots, it is well-understood that the CPhase/CZ gates can be realized with high fidelity via accurately controlled DC evolutions \cite{Meunier2011Efficient,Russ2018Highfidelity,Qi2024Spin}.
In this section, we develop a theoretical extension for this important gate class,  investigating what are the possible multi-qubit gates that can be similarly implemented on a general array of quantum dots with DC control.

\subsection{The qubit-frame map}
Quantum gates are associated with unitary maps in the ``qubit frame''.  It is thus necessary to first clarify relevant concepts about this frame. In comparison to a global rotating frame that matches the external driving field, 
the qubit frame is a direct product of many locally rotating frames associated with the qubits \cite{Vandersypen2004NMR}. 
We recall that the Hamiltonian governing a qubit system can always be split as $H=H_0+H_\mr{I}$,
where $H_0$ is responsible for the proper definition of the qubits and remains static within the time span of interest; $H_\mr{I}$ includes all the external control signals and internal interactions necessary for manipulating the qubits. 
Setting the initial instance of evolution as $0$,  then $H_0$ 
induces the unitary transform 
$\ee^{-\ii  \tau H_0}$ to the lab-frame states after temporal duration $\tau$. 
The states and observables in the qubit frame are defined by $X_\mr{q}\equiv\ee^{\ii  \tau H_0} X_\mr{lab}\ee^{-\ii  \tau H_0}$. 
The time evolution operator $\Uqb(\tau)$ for qubit-frame states is generated by the interaction-picture Hamiltonian 
$H_\mr{q}(\tau)=\ee^{\ii  \tau H_0} H_\mr{I} \ee^{-\ii  \tau H_0}$. Such $\Uqb(\tau)$ can be expressed as a Dyson series \cite{Sakurai2020Modern}, but explicit time-dependence in $H_\mr{q}(\tau)$  often makes exact time integration difficult. 

To study the set of multi-qubit DC gates, we focus on a single time evolution segment with static exchange coupling, i.e., $H_\mr{I}=H_\mathrm{ex}$ independent of time. This allows us to directly express the time evolution as
\begin{equation}\label{eq:Ut}
\Uqb(\tau) = \ee^{+\ii \tau H_0} \ee^{-\ii \tau (H_0+H_\mathrm{ex})}.
\end{equation}
For this study, we are concerned with whether $\Uqb(\tau)$ can faithfully represent a useful quantum gate at a certain time $\tau$.
Using the Magnus expansion \cite{Blanes2009Magnus},
an effective Hamiltonian can be derived for $\Uqb$ that works well within the short time limit $\tau \norm{H}\ll 1$. However this approach is not viable for quantum gate problems with the timescale 
$\tau\sim\norm{H}^{-1}$.  For accurate description of the long-time behavior, it is necessary to diagonalize the matrix exponents.  This can also be difficult analytically. Exact results are only known to us for the basis case of a 2-qubit system. To gain insights of the qubit-frame map for large systems, we thus resort to approximate treatments combined with error analysis.

For common spin-qubit systems fabricated in current laboratories, the Zeeman splitting energies of quantum dots are on the order of $1{-}10$~GHz and the interdot exchange energies are of the order $10{-}100$~MHz \cite{Noiri2022Fast}.  We assume that our hypothetical quantum chip follows similar energy scales. According to \Eqref{eq:Heff}, this energy hierarchy implies $\norm{H_0}\gg \norm{H_\mr{ex}}$ and hence permits a perturbative treatment for the time-evolution.
Apparently, $H_0$ is already diagonal under the computational basis, with the eigenenergy 
\begin{equation}\label{eq:H0Energy}
E_n = \sum_j \frac{1}{2} \sig(\sigma_{n_j} ) \, \Ez_j,
\end{equation} 
for the eigenstate $\ket{n}$. 
Let us denote the eigenstates and eigenenergies of $H$ by  $\{\ket{\wt n}\}$ and $\{\wt E_n\}$ respectively.  Then the matrix exponentials in \Eqref{eq:Ut} can be carried out in the relevant eigenstate basis. 
In particular, its diagonal elements are found by  
\begin{equation}\label{eq:Ut-diag}
 \bra{n}{\Uqb(\tau)}\ket{n} = \sum_{m}\, \abs{\braket{n}{\wt m}}^2 \, \ee^{-\ii \tau(\wt E_m - E_n)}.
\end{equation}
Under the perturbative assumption, eigenenergies  and eigenstates of the full Hamiltonian $H=H_0+H_\mr{ex}$ are slightly shifted from that of $H_0$ due to the presence of a small $H_\mr{ex}$ term. As a result, the sum in \Eqref{eq:Ut-diag} can be split into a major term of magnitude  $\abs{\braket{n}{\wt n}}^2= 1-O(J^2)$, and a sum many of minor terms on the order of
$\abs{\braket{n}{\wt m}}^2=O(J^2),\ m\neq n$. It can be also shown that the off-diagonal terms of $\Uqb$ are on the order of $O(J^2)$.
Expanding $\Uqb$ by the coupling strength $J$, 
we obtain the first-order map
\begin{equation}\label{eq:Uideal}
\begin{aligned}
 \Uideal(\tau) &= \sum_n \ee^{-\ii \tau \delta E_n^{(1)}} 
\dyad{n} 
\equiv \ee^{\ii\tau\Lambda},
\end{aligned} 
\end{equation}
where 
$\{\delta E^{(1)}_n \equiv\bra{n}H_{\mr{ex}}\ket{n}\}$ are the first-order energy corrections, which are further used to define 
 the generator $\Lambda$. 
We refer $\Uideal$ as the \emph{ideal} map, which is to be later identified as a useful quantum gate. By the perturbative expansion of $\Uqb$, we treat DC quantum gates as first-order dynamical effects of the exchange coupling. In comparison, the zeroth-order map $\ee^{-\ii \tau H_0}$ defines the qubit frame. While all the second or higher-order terms can be attributed as \emph{coherent errors} in the gate implementation.  

We should note that despite $\Uideal$ being the leading order term of the perturbative expansion, there is no guarantee that it approximates the {actual} map $\Uqb$ in general.
This is because the first-order energy corrections involved in \Eqref{eq:Uideal} are derived from non-degenerate perturbation theory. If a pair of energy levels in $H_0$ are very close, which is in fact quite probable for a large array of quantum dots, we expect the approximation 
$\Uqb \approx \Uideal$ to fail (even in the small $J$ limit). 
Hence our following discussions necessarily split into two related parts. In \ref{sec:intriG-class}, we study the algebraic structures and properties of multi-qubit gates achievable by $\Uideal$. In \ref{sec:fidelity}, we investigate the validity of the ideal map approximation and prove that the coherent gate errors are well-bounded under practical conditions with suitable chip designs.

\subsection{The DC gate family }\label{sec:intriG-class}
\subsubsection{Array and bond vectors}
According to \Eqref{eq:Uideal}, under the computational basis,  $\Uideal$ contains only simple phase factors on the diagonal entries 
and its generator can be defined by $\Lambda=-\diag(H_\mr{ex})$. Using \Eqref{eq:Heff} and \Eqref{eq:ent-state},
we obtain the decomposition 
\begin{equation}\label{eq:arrayvec}
\Lambda
=\bigoplus_{\mathclap{{w=\bond{j,k}}}} \Lambda_w
= \bigoplus_{w}
\begin{pmatrix}
  S_w &&&\\
& T_w &&\\
&&T_w &\\
&&& S_w
\end{pmatrix}_{\mathrlap{(w-\mathrm{subspace})}},
\end{equation}
where the Kronecker sum ($\oplus$) is defined for two operators $x\in {A}$ and $y\in {B}$ in possibly different linear spaces by 
$x \oplus y = x \otimes \mathbb{I}_{ B\backslash  A} +  \mathbb{I}_{ A \backslash  B}\otimes y$, where $\mathbb{I}_{B\backslash A}$ is the  identity operator supported on the subspace $B\backslash A=\{x|x\in B, x\notin A\}$. For example, consider a triple-dot system where only direct exchange coupling between qubit 1-2 and qubit 2-3 are allowed. 
Both $\Lambda_{(1,2)}$ and $\Lambda_{(2,3)}$ are defined in their respective subspaces. With the Kronecker sum, they combine into the 8-dimensional  
$\Lambda=\Lambda_{(1,2)} \oplus \Lambda_{(2,3)}=\Lambda_{(1,2)} \otimes \begin{psmallmatrix}
  1 &\\
& 1
\end{psmallmatrix}_{(3)} + \begin{psmallmatrix}
  1 &\\
& 1
\end{psmallmatrix}_{(1)} \otimes \Lambda_{(2,3)}$.

As $ \Lambda$ depends on all connecting bonds in the array while $ \Lambda_w$ depends only on a particular bond $w$, 
we refer the former as the array generator and later as the bond generator. 
In \Eqref{eq:arrayvec}, each bond generator is explicitly represented as a diagonal matrix under the computational basis 
$\{\ket{\uparrow\uparrow}_{jk},\ket{\uparrow\downarrow}_{jk},\ket{\downarrow\uparrow}_{jk},\ket{\downarrow\downarrow}_{jk}\}$, with the spin-conserved and the spin-flipped tunneling energy 
\begin{align}\label{eq:tunnel-energy}
S_w \equiv \frac{J_w}{2} \frac{\abs*{s_w}^2}{\abs{s_w}^2+\abs{t_w}^2}, \quad 
T_w \equiv \frac{J_w}{2} \frac{\abs*{t_w}^2}{\abs{s_w}^2+\abs{t_w}^2}, 
\end{align}
with $\tr(\Lambda_w)=2S_w+2T_w=J_w$ representing the exchange energy 
of the bond.

For simplicity,  we shall address $\Lambda$ and $\Lambda_w$ by the 
$2^N$-dimensional \emph{array vector} $\bm \Lambda$ 
and the 4-dimensional \emph{bond vectors} $\bm \Lambda_w$ defined by the diagonal elements of  the corresponding matrix representations. 
These vectors are denoted with bold fonts and should be easily distinguishable with their operator counterparts by the context.
The components of $\bm \Lambda$  are summations of various 
 $S_w$ and $T_w$ terms.  Since different bonds can overlap, the exact expression of the array vector depends on the ordering of qubits and the topology of the array, and thus can be quite involved.
We can nevertheless take note that the array vector
is always reflectively symmetric,
\begin{equation}\label{eq:reflective}
 {\bm\Lambda} = \overleftarrow{\bm \Lambda}  \equiv 
\bigl(\, {\bm \lambda} , \overleftarrow{\bm \lambda}\,\bigr),
\end{equation}
where the left over-arrow represents reversing the element order of a given vector. For example, if $\bm a= (a_1,a_2,a_3,a_4)$ then 
$ \overleftarrow{\bm a}=(a_4,a_3,a_2,a_1)$. A proof of this property is given in Appendix~\ref{app:refsym}. This reflectively symmetry implies that there are at most one-half independent entries in
 ${\bm\Lambda} $, defined by the \emph{reduced} array vector $\bm\lambda$ in \Eqref{eq:reflective}.

\begin{figure}[htbp]
  \centering
  \includegraphics[width=10cm]{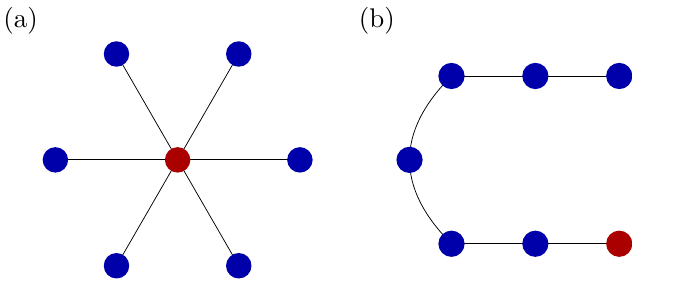}
  \caption{Two basic types of array topology: (a) stellar topology, (b) chain topology. The red dot in each graph marks the first qubit in the Hilbert space for the reduced array vector formula in \Eqref{eq:lambda-stellar} and \Eqref{eq:lambda-linear}.
We also note that the array topology depends only one the way how the dots are directly coupled by exchange interaction, not by their relative position or distance.}
  \label{fig:arrtopo}
\end{figure}

To demonstrate the geometric dependence of the array vector, let us consider an array of stellar topology. As shown in \Figref{fig:arrtopo}(a), the dot cluster  in stellar topology resembles a star with a center dots connecting to all the rest dots. Taking the central dot as the first qubit in the Hilbert space,  the reduced array vector can be worked out as 
\begin{equation}\label{eq:lambda-stellar}
 \bm \lambda^{(\text{star})} =\bigoplus_{j\ge 2} (S_{1j},T_{1j}),
\end{equation}
with $j\ge2$ labeling the dots on the ends.  In comparison, let us also examine another type of array topology---the chain topology. As shown in \Figref{fig:arrtopo}(b), in such configuration, all qubits are sequentially connected with into 1-dimension string. 
It can be shown that the reduced array vector $\bm \lambda^{(N)}$ of an $N$-dot chain ($N\ge 3$) satisfy the following recurrence relation 
\begin{equation}\label{eq:lambda-linear}
 \bm \lambda^{(N)}= \bm \lambda^{(N-1)}\otimes(1,1) +\bm \lambda^{(N-2)}\otimes(S_{N-2},T_{N-2},T_{N-2},S_{N-2}), 
\end{equation}
and can be deducted recursively from $\bm\lambda^{(2)}=(S_{1},T_{1})$ and $\bm\lambda^{(1)}=1$, 
where $S_n$ and $T_n$ stand for the spin-flipped and conserved tunneling energy for the bond connecting the $n$th dot to the $(n+1)$ dot in the chain.

\subsubsection{Phase gauge}
To relate the time evolution map $\Uideal(\tau)$ with a useful quantum gate, it is customary to include the effects of additional global phase factor and local phase gates. We refer these combined phase degrees of freedom as the phase gauge. 
First, two unitary maps differing by a global phase factor $\ee^{\ii \phi_\mr{g} }$ are completely equivalent. This equivalence constitutes the global phase gauge. 
More importantly, two unitary gates are also considered equivalent if they differ by local phase gates 
$Z(\phi)=\ee^{-\ii (\phi/2) \sigma^\mr{Z}}$ on individual qubits. 
The rationale behind such local phase freedom is that single-qubit $Z$-axis rotations can be implemented by the so-called virtual-$Z$ gates \cite{Vandersypen2004NMR,McKay2017Efficient}. If the interqubit coupling Hamiltonian commute with
Z on each qubit [such as array generator in ~\Eqref{eq:arrayvec}], these phase rotations can be naturally combined and eliminated. On the other hand, for circuit containing single-qubit gates not commuting with $Z(\phi)$, one can use various circuit compilation techniques, such as gate permutation and pulse-level engineering to eliminate physical $Z$ rotations. Interested readers may refer to, for example, Ref. \cite{mckay_efficient_2017} and \cite{Long2025Virtual} for recent discussions on this manner. In this paper, we assume that such single-qubit $Z(\phi)$ gates is a freedom that come at no cost of fidelity nor operation time.
Such phase freedom is implied in the original proposal of the CPhase gate \cite{Meunier2011Efficient}, and further exploited in fidelity optimizations \cite{Russ2018Highfidelity,Wu2023Hamiltonian}. It is natural to extend this freedom in phase gauge for multi-qubit gates. 

Combing the global phase freedom with the direct product of local phase gates for all the qubits,
we can have the full phase gauge transform,
\begin{equation}\label{eq:Zfree}
\begin{aligned}
Z(\bm\phi)& = \ee^{\ii\phi_\mr{g}}\, \Bigl[
\bigotimes\nolimits_j\! Z_j(\phi_j) \Bigr]\equiv \ee^{\ii \Phi}.
\end{aligned}
\end{equation}
Here a Lie algebra exponent $\Phi$ is naturally defined. We can further introduce the gauge vector by its diagonal matrix elements in the computational basis,
\begin{equation}\label{eq:gauge-vec}
\bm\Phi=\diag(\Phi)=\phi_\mr{g}+\bigoplus_j (0,\phi_j).
\end{equation}
We now formally define the multi-qubit DC gate family as the set of multi-qubit gates that can be theoretically achieved on a spin qubit array by an ideal map together with a phase gauge transform, $\{G= Z(\bv\phi)\, \Uideal(\tau)\}$.
 As both $\Uideal(\tau)$ and $Z(\bv\phi)$ are diagonally represented in the computation basis, $G$ should also be diagonal in the computational space. 
Unitarity of $G$ requires that such $G = \ee^{\ii \Theta_\mr{G}}$,
where the diagonal ``gate vector''  ${\bm \Theta}_\mr{G}=\diag(\Theta_\mr{G})$ contains only real elements. Focusing only on the exponent parts, we can obtain an equivalent relation for multi-qubit DC gates,
\begin{equation}\label{eq:vec-relation}
 \bm\Theta_\mr{G}  = \bm\Phi+ \tau\bm\Lambda \mod 2\pi ,
\end{equation}
where the $2\pi$ modulus applies to all the vector components.

\subsubsection{Multi-qubit controll-phase gates}\label{sec:generalCPhase} 
Given the CPhase gate is directly implementable on a double-dot system, it is natural to ask whether its multi-qubit extensions are equally applicable on an array of quantum dots. 
Broadly speaking, such a multi-qubit gate can induce
 conditional $Z$-axis rotations to the target qubits provided that the control qubits are all $\ket{1}$'s (here we take $\ket{0}\equiv\ket{\uparrow}$ and $\ket{1}\equiv\ket{\downarrow}$), and does nothing otherwise. 

Without loss of generality, we can  fix the \emph{first} qubit as the control qubit. This choice implies that the gate vector can be split as
${\bm \Theta}_\mr{G}=(\bm 0,\bm\theta_\mr{G})$,
where the zero vector $\bm 0$ has the same length as the \emph{reduced} gate vector $\bm\theta_\mr{G}$. 
By half-splitting all vectors
in \Eqref{eq:vec-relation} and taking advantage of the reflective property [\Eqref{eq:reflective}], we obtain two gate composition rules,
\begin{align}
  \bm \theta_\mr{G} &= \phi_{1}+\bigoplus_{j\ge 2} (-\phi_j,\phi_j) 
\mod 2\pi, \label{eq:parity-rule} \\
 \tau \bm\lambda &= -\phi_\mr{g} - \bigoplus_{j\ge 2} (0,\phi_j)  \mod 2\pi. \label{eq:dynamics-rule}
\end{align}
where $\phi_1$ is the phase gauge for the control qubit, with $j\ge 2$ labeling the rest qubits.
Since \Eqref{eq:parity-rule} depends only on the target gate and the number of qubits, we refer to it as the ``parity rule'', which restricts the accessibility of a given gate $G$.  Meanwhile \Eqref{eq:dynamics-rule} connects the local phase gauge, the array vector and the evolution time, hence is referred as the ``dynamics rule'' in followings. 

Consider the basic example of a 2-qubit CPhase gate,
defined as $\CZ_\theta =\diag(1,1,1,\ee^{\ii \theta})$ in the computational basis and is recognized as the $\CZ$ gate if the target phase shift $\theta=\pi$. 
It can be verified that the parity rule [\Eqref{eq:parity-rule}] is fully invertible for any $\theta\in (0,2\pi)$. Substituting the phase gauge solutions
$(\phi_1=\phi_2=\theta/2-n\pi,\ n\in \mathbb{Z})$ and the double-dot vector $\bm\lambda=(S,T)$ into the \Eqref{eq:dynamics-rule}, we reach the target CPhase gate at 
\begin{equation}\label{eq:tau_tqg}
 \tau=\frac{-\theta+2n\pi}{2(T-S)}, \quad \text{for } n\in \mathbb{Z}, \text{ and } \tau>0.
\end{equation}
If we focus on the minimal positive $\tau$, then the integer $n$ for the phase gauge can be take to be $n=\frac{1}{2}(1+\operatorname{sgn}(T-S))$. This recovers a result of Ref.~\cite{Qi2024Spin} for the $\theta=\pi$ case.

\smallskip
For an $N$-qubit array, $\bm \theta_\mr{G}$ is $2^{N-1}$ dimensional but there are only $N$ local phases to vary, hence a large number of gates are naturally prohibited by the parity rule. 
In particular,  the parity rule suggests the following restriction.

\smallskip
\noindent
\textbf{Proposition} \textit{For multi-qubit DC gates, the number of control qubits cannot exceed 1.}

\smallskip
\noindent\textit{Proof.} 
Assume that a multi-qubit CPhase gate $G$ with at least two control qubits exists.
Without loss of generality, let us set both the first and the second qubit as control qubits. This implies that $\bm\theta_G=(\bm 0', \bm \theta_G')$, where $\bm 0'$ is a zero-vector of length $2^{N-2}$.  Let us split both sides of the parity rule [\Eqref{eq:parity-rule}] into equal halves and substitute into $\bm\theta_G$,
yielding
\begin{equation}
\left\{
 \begin{aligned}
 & \phi_{2}-\phi_{1} + \bigoplus\nolimits_{j\ge 3} (-\phi_j,\phi_j) =\bm 0'\\
 &\phi_{2}+\phi_{1} + \bigoplus\nolimits_{j\ge 3} (-\phi_j,\phi_j) =\bm \theta_G'
 \end{aligned} 
\right. \mod 2\pi.
\end{equation}
Solving this condition, we find 
$\bm \theta_G'=2\phi_{2}$, which is a constant vector (up to $2\pi$ modulus). Such solution in turn suggests that $G$ is a CPhase between the first two qubits instead of a multi-qubit gate. Hence it is contradictory to our assumption. $\qedsymbol$

Given the above, we only need to consider multi-qubit controlled-phase gates with one control qubit and multiple target qubits, in the form of 
$\mathrm{CZ_{\theta_2}Z_{\theta_3}\cdots Z_{\theta_{N}}}$.
For briefly, we refer such gates as the multi-target controlled-phase (MTCP) gates. 
Further analysis of the dynamics rule [\Eqref{eq:dynamics-rule}] reveals that such a gate is applicable if and only if the dots form the stellar topology, whose reduced array vector $\bm\lambda$ is of the form in \Eqref{eq:lambda-stellar}.
Assuming $0\le\theta_j<2\pi$ for all $j$, we can work out from \Eqref{eq:parity-rule} the local phase factors as
\begin{equation}\label{eq:cphase-gauge}
\begin{aligned}
\phi_1 &= \sum_{j\ge2} \phi_j \mod 2\pi,\\
\text{and}\quad 
\phi_j &=\frac{1}{2}\theta_j - n_j \pi \quad (n_j \in \mathbb{Z}),\ \text{for } 2\le j \le N. 
\end{aligned}
\end{equation}
Substituting $\bm \lambda=\bm \lambda^{(\text{star})}$ from \Eqref{eq:lambda-stellar}, we can solve \Eqref{eq:dynamics-rule} by $\phi_\mr{g}=-\tau \sum_j  S_{1j}$ and 
$\phi_j = -\tau(T_{1j}-S_{1j})$. Combined with \Eqref{eq:cphase-gauge}, this suggests the following condition for the bonds
\begin{equation}\label{eq:cphase-dynamics}
 \tau(T_{1j}-S_{1j})= -\frac{1}{2}\theta_{j} + n_j\pi,  \quad (n_j \in \mathbb{Z}).
\end{equation}
We remark that \Eqref{eq:cphase-gauge} suggests how to \emph{decode} the first-order dynamical map as a MTCP gate by applying proper local phase corrections, while \Eqref{eq:cphase-dynamics} sets constraints on the array connectivity and gate evolution time.

A general MTCP gates cannot be implemented in the chain topology, with array vector specified in \Eqref{eq:lambda-linear}.  It is however interesting to note one exception: 
for a homogeneous chain with equal coupling (namely $S_n=S,\ T_n=T$ with $T\neq S$ for all bonds), a $\pi$-phase can accumulate on both ends $\phi_1=\phi_N=\pi$ while coherently cancels out for other qubits
 $\phi_2=\phi_3=\cdots\phi_{N-1}=0$.  In such case, the resulting gate is simply 
\begin{equation}
 \Uideal(\tau =\pi/\abs{T-S} ) = Z_1\otimes I_2\otimes I_3\otimes\cdots \otimes I_{N-1}\otimes Z_{N}.
\end{equation}
We remark that the superexchange oscillations observed for boundary states of a spin chain are manifestation of this gate \cite{Qiao2020Coherent,Qiao2021LongDistance}. While this gate seems  non-entangling, in any intermediate time  $0<\tau<\pi/\abs{T-S}$ the states can be entangled across the chain. When measuring the spin probability at
the edge dots, one obtain an correlated oscillation of their spin polarization. 
Such gate might be trivial from the quantum computational perspective, as it is just joint single-qubit $Z$. But the underlying physics is quite interesting. 
The final gate acts on remote edge modes and does not depend on the shape and number of qubits involved in the connecting path. Hence it has the property of being topologically invariant and can be potentially used in error correction codes. 

\subsubsection{Decomposing general multi-qubit gates}\label{sec:decompose}
For a given array of quantum dots, there are many different possible ways for making interdot connections.  The MTCP gates associated with stellar topology only constitute a small subset of the multi-qubit DC gate family. However, our understanding of the later can be greatly simplified by the following theorem. 

\medskip
\noindent
\textbf{Theorem} \textit{All multi-qubit DC gates of a quantum-dot array are equivalent to the simultaneous product of 2-qubit controlled-phase gates up to some 
local phase gauges determined by the array connectivity.}

\smallskip
\noindent\textit{Proof.} 
The goal is to prove the existence of an appropriate phase gauge $\bm\phi=(\phi_\mr{g},\phi_1,\phi_2,\cdots)$ such that the following equality holds,
\begin{equation}\label{eq:Uideal-decompose}
Z(\bm\phi)\Uideal(\tau) {=} \prod_{\mathclap{w=\langle{j,k}\rangle}}\, \mr{C}_j\mr{Z}_k( \theta_w),
\end{equation}
where for each $\mr{CZ}$ gate in the above product a tensor-product with $\mathbb{I}_{w\perp}$, namely the identity operator on the orthogonal subspace of bond-$w$, is assumed.  
As a side note, 
the subscripts $(j,k)$ for the above $\mr{C}_j\mr{Z}_k$ can be swapped as there is no distinction between the control and the target qubit for a 2-qubit CPhase gate. It is only the relative phase shift $\theta_w$ between the two qubits that matters. 

\begin{figure}[thbp]
 \centering
 \includegraphics[width=10cm]{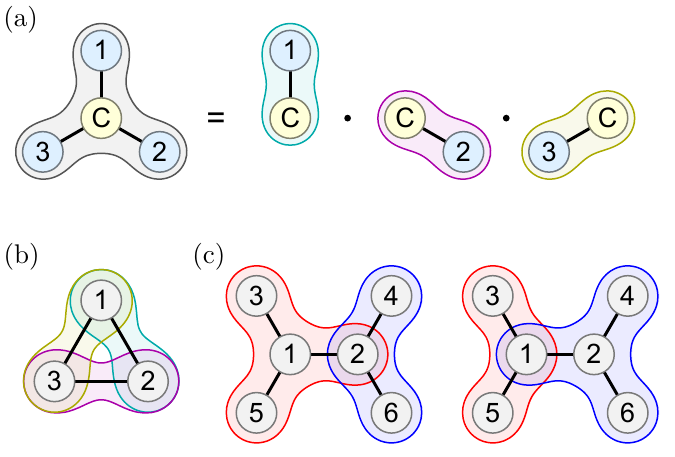}
 \caption{Decomposition of multi-qubit DC gates based on the interdot connectivity.
(a) The DC gate for a 4-qubit system in stellar topology is an multi-qubit CPhase gate, which can be decomposed as the product of 3 regular CPhase gates.
(b) The DC gate of a 3-qubit ring does not contain a control qubit, as the 3 qubits are totally symmetric to each other. 
(c) Two different ways of decomposing the same DC gate for a 6-qubits array as product of 2 MTCP gates.  The resulting local phase corrections are independent  of particular choice of decomposition. 
}
 \label{fig:decompose}
\end{figure}

Since \Eqref{eq:Uideal-decompose} already holds for double-dot systems, we can assume it also holds for arrays with up-to $N\ (N\ge2)$ dots and consider proving it for an array of $N+1$ dots. We note that by construction all bond generators $\Lambda_w$ of an array commute against each other.  Hence the time-evolution operator for the extended array can be reduced as 
\begin{equation}
\Uideal(\tau)
= \prod_{\mathclap{\substack{w=\bond{j,k}\\1\le j,k\le N+1}}} 
\,\ee^{\ii\tau {\Lambda}_{w}} 
= \prod_{\mathclap{\substack{w=\bond{j,k}\\1\le j,k\le N}}} 
\,\ee^{\ii\tau {\Lambda}_{w}} 
\, \prod_{j=1}^N \ee^{\ii\tau {\Lambda}_{(j,N+1)}},
\end{equation}
where a tensor product with orthogonal identity is implied for each term in the product. By assumption, the first part of the right-hand-side is equivalent to a CPhase-product under a suitable phase gauge.
Meanwhile, we recognize the second part as the time-evolution operator for a stellar array, where the first $N$ dots are exclusively connected to the last dot. Hence it carries out a MTCP gate as 
discussed in Sec.~\ref{sec:generalCPhase}. 
We note that a MTCP gate can be viewed as the product of multiple CPhase gates sharing a common qubit, as illustrated in \Figref{fig:decompose}(a) for a 4-qubit example. For a MTCP gate, the control/target qubits naturally stand out as the cluster center/ends, hence the control-target symmetry for the 2-qubit case is broken.
Applying the induction assumption and the solutions in \Eqref{eq:cphase-gauge} and \Eqref{eq:cphase-dynamics}, we find that the phase gauges for the $N$-array and the $(N+1)$-array are related by,  
\begin{equation}\label{eq:gauge-recursion}
\left\{
\begin{aligned}
&\phi_\mr{g}^{[N+1]} =\phi_\mr{g}^{[N]} +\sum_{\mathclap{j\le N}} \tau S_{(j,N+1)},\\
&\begin{aligned}[t]
  \phi_j^{[N+1]}&=\phi_j^{[N]}-\tau\! \left[T_{(j,N+1)}-S_{(j,N+1)}\right]\\
&=\phi_j^{[N]}+\frac{1}{2} \theta_{(j,N+1)} -n_j \pi
, \quad \ n_j \in \mathbb{Z},\ 1\le j \le N,
\end{aligned}\\
&\phi_{N+1}^{[N+1]} =  \sum_{j=1}^N \,\Bigl[\phi_j^{[N+1]} -\phi_j^{[N]}\Bigr],
\end{aligned}\right.
\end{equation}
where the superscripts in square brackets label the array size.
This set of relations can be recursively applied to determine the phase gauge for the extended array.  Hence we have proven \Eqref{eq:Uideal-decompose} by mathematical induction.  \qed 

For a given quantum-dot array of with known connectivity, we can derive from \Eqref{eq:gauge-recursion} the following simplified expressions for the phase gauge
\begin{align}
\phi_\mr{g}&= -\tau \sum_{w} S_w \mod 2\pi,\label{eq:global-phase} \\
  \phi_j &= -\tau \sum_{w:\,{j\in w}} (T_w-S_w) \mod 2\pi.\label{eq:local-phase} 
\end{align}
In particular, we obtain from \Eqref{eq:local-phase} an intuitive understanding that the phase correction for a particular quantum dot is determined by the combined phase shifts from all its connecting bonds. This also explains the local phase relation $\phi_\mathsf{C}=\sum_j \phi_j$  [\Eqref{eq:cphase-gauge}] for a MTCP gate, where the local phase of the control qubit is the sum of that of target qubits. 
In the gauge specified by \Eqref{eq:global-phase} and \Eqref{eq:local-phase}, $\Uideal(\tau)$ can be decomposed as a product of 2-qubit CPhase gates as in \Eqref{eq:Uideal-decompose}, with 
the phase shift
\begin{equation}\label{eq:mqgatetime}
 \theta_w = -2\tau (T_w-S_w) \mod 2\pi,
\end{equation} 
for each bond $w$ in the array.
We note that this condition should be simultaneously satisfied by all the bonds. Hence if the target gate is determined \emph{a priori}, it imposes restrictions on the \emph{effective} bond strengths, defined by the difference $T_w-S_w$ for each bond $w$.

This decomposition theorem for multi-qubit DC gates, combined with the phase gauge solutions, provides several key insights into this gate family. 
It suggests that despite all the possible interdot connections, the algebraic structure of all multi-qubit DC gates on an array is locally equivalent to the simultaneous product of 2-qubit CPhase gates, with the local phase corrections efficiently calculable from the array connectivity (topology + bond strength). The MTCP gates considered in Sec.~\ref{sec:generalCPhase} are special cases of this  theorem. Furthermore, this theorem provides a systematic approach to construct various multi-qubit DC gates apart from the MTCP gates. 
For example, consider the triple-dot ring in  \Figref{fig:decompose}(b). No particular dot can serve as the control qubit due to its symmetric arrangement, therefore the 3-qubit DC gate on such system does not belong to the MTCP class. Meanwhile, for large arrays with 
complicated interdot connections, the associated multi-qubit DC gates can be understood by decomposing the full array into multiple stellar-connected subgroups, each of which corresponds to a MTCP gate to be simultaneously implemented. 
There can usually exist different possible ways of decomposition, which are just equivalent interpretations of the same gate. Such is demonstrated in \Figref{fig:decompose}(c) for a cluster of 6 qubits. In particular, the local phase corrections should be invariant for different decompositions.  For the example in \Figref{fig:decompose}(c),
assuming that all bonds are homogeneous and satisfy the condition $(T-S)\tau=\pi/2$, we calculate $\phi_1=\phi_2=3\pi/2$ and $\phi_3=\phi_4=\phi_5=\phi_6=\pi/2$ from both ways of decomposition.  
Finally, we remark that while the multi-qubit gates considered here can be equivalently implemented with a product of 2-qubit CPhase gates, so is any multi-qubit unitary transformation (since CPhase is a universal gate). Multi-qubit DC gates are of significance by themselves as an integral unit for quantum computation.  
We will construct more examples of such multi-qubit DC gates and discuss their advantages over regular two-qubit gates in Sec.~\ref{sec:examples}.

\subsection{Estimating and optimizing gate errors}\label{sec:fidelity}
We have hitherto replaced the qubit-frame map $\Uqb$ with the ``ideal'' map 
$\Uideal$ and investigated what can be achievable with the latter.
However, the conditions for $\Uqb\approx\Uideal$  is more subtle than just requiring 
$\norm{H_0}\gg \norm{H_{\rm{ex}}}$. 
In this section, we examine in detail how much error is brought by making such approximation and look for potential ways to reduce these errors. 

\subsubsection{Gate infidelity upperbound}
The quality of an implementation of a particular quantum gate is usually characterized with the average gate fidelity \cite{nielsen2002simple}.  
Here, we focus on estimating the coherent gate fidelity, where the ideal gate $Z(\bm\phi)\Uideal$ is implemented by $Z(\bm\phi)\Uqb$. Since both are unitary operators of the same dimensionality, the gate fidelity can be calculated by 
\begin{equation}\label{eq:fidelity}
\Fid(\Uideal,\Uqb)= \frac{d+\abs{\tr(\Uideal \Uqb^\dagger)}^2}{d(d+1)},
\end{equation}
where $d=2^N$ is the dimension of the system. A helpful simplification is brought by the fact that $\Uideal$ is diagonal in the computational basis. Hence to calculate the gate fidelity, one only requires the diagonal entries of  $\Uqb$ defined in \Eqref{eq:Ut-diag}. 
Using a careful combination of inequalities, we can derive (in Appendix~\ref{app:gatefid}) the following upper-bound for the gate infidelity $\InF\equiv 1-\Fid$,
\begin{align}\label{eq:inf-bound}
  \InF &\le  \frac{4}{d+1}\sum_n (1-\abs{\braket{\wt n}{n}}^2)
+ \frac{1}{d+1}\sum_{n} [\tau\,\delta E_n^{(2+)}]^2 \notag \\
&=\frac{d}{d+1} (4e_\mr{S}+e_\mr{P}). 
\end{align}
Here we single out two size-independent factors for the gate infidelity: the ``state error'' $e_\mr{S}\equiv 1/d\sum_n(1-\abs{\braket{\wt n}{n}}^2)$ from the non-unital overlap between the perturbed and original eigenstates, and
the ``phase error'' $e_\mr{P}\equiv1/d\sum_n [\tau\,\delta E_n^{(2+)}]^2$ due to phase shifts accumulated from higher-order energy corrections $\delta E_n^{(2+)}\equiv\wt E_n - E_n -\delta E_n^{(1)}$. To evaluate these error terms, we carry out second-order perturbations to find,
\begin{equation}\label{eq:coherrs}
\begin{aligned}
e_\mr{S}\approx e_\mr{S}^{(2)}&= \frac{1}{d}\sum_n\sum_{m \neq n}
\frac{\abs[\big]{ \bra{n} H_\mr{ex} \ket{m} }^2 }{(E_{n}-E_{m})^2}, \\
e_\mr{P}\approx e_\mr{P}^{(2)}
&= \frac{\tau^2 }{d} \sum_n \Bigl(\sum_{m\neq n} \frac{\abs[\big]{ \bra{n} H_\mr{ex} \ket{m} }^2}{E_n-E_m} \Bigr)^2,
\end{aligned}
\end{equation}
where the superscripts indicate the perturbative order. 
Since the matrix elements of $H_\mr{ex}$ are linear in $J$, while according to \Eqref{eq:mqgatetime} the evolution time $\tau\propto J^{-1}$, we conclude that $e_\mr{S}$ and 
$e_\mr{P}$ are both of order $O(J^2)$. 
It is worth noting that \Eqref{eq:coherrs} are derived with simple DC control in mind, where the exchange coupling is a rectangular function signal in time. With the help of adiabatic pulse shaping, these errors can be further reduced  to 
the  square of the spectral power of the pulse at the detuning frequency  \cite{Rimbach-Russ2023Simple,Polat2025Pulse}. Here, simple DC control is sufficient for achieving a theoretical upper-bound on the error rates and advanced pulse shaping and optimization schemes are not considered.

The above analysis verify the perturbative hierarchy that the ideal map $\Uideal$ is of first order and coherent errors are of second order in the series expansion of $\Uqb$. However, this hierarchy does not guarantee $\Uqb\approx\Uideal$ in general. 
As revealed from \Eqsref{eq:coherrs}, the entire perturbative treatment could breakdown should a pair of unperturbed energy levels became nearly degenerate. Let us consider an array of $N$ quantum dots with randomly distributed Zeeman energies. 
According to \Eqref{eq:H0Energy}, there are in total different  $2^N$ energy levels within an energy range growing linearly with $N$. As a result, the minimal energy level detuning $\langle\min_{n\neq m}\abs{E_n-E_m}\rangle$ is expected to decrease by $e^{-N}$. 
Hence accidental degeneracy is inevitable as the system scales up. 
For double-dot systems, such energy degeneracy can be artificially lifted, e.g., by applying a large magnetic field gradient across the dots \cite{Kawakami2016Gate}.
Engineering a similar non-degenerate condition is however   impractical for large arrays.

While energy degeneracy seems detrimental for multi-qubit gates, closer inspection of \Eqsref{eq:coherrs} suggests that full non-degeneracy is in fact quite unnecessary. 
This is because for a large number of basis state pairs $(n,m)$, the matrix elements $\bra{n} H_\mr{ex} \ket{m}$ vanish and hence the corresponding terms are naturally excluded from the sums in \Eqsref{eq:coherrs}. 
By construction, the ``selection rule'' for non-zero matrix elements can by formulated 
as follows: if we expand the basis states $\ket{n}$ and $\ket{m}$ as binary strings [as in \Eqref{eq:comp-state}] and compare them bit-wise, then $\bra{n} H_\mr{ex} \ket{m}\neq 0$ only when the Hamming distance (i.e., number of differing components) between $\ket{n}$ and $\ket{m}$ is less-or-equals to 2.
Consequently, the second order error terms can be further decomposed into finer contributions from each bond,
\begin{align}\label{eq:err-decompose}
 e^{(2)}_\mr{S}=\sum_w e^{(2)}_{\mr{S},w}, \quad 
 0<e_\mr{P}^{(2)}\lesssim 2\sum_w e^{(2)}_{\mr{P},w},
\end{align}
where the secondary subscripts for $e_\mr{S}$ and $e_{P}$ stand for the corresponding error contributions from a particular bond. The ``$\lesssim$'' sign for $e_\mr{P}$ in \Eqref{eq:err-decompose} holds well for typical systems where $\Ez\gg J$.  Detailed proofs of these results are given in Appendix~\ref{app:errdecomp}.
For a particular bond $w=\bond{j,k}$, we can explicitly derive the leading-order error expressions,
\begin{equation}\label{eq:err2dots}
  \begin{aligned}
e^{(2)}_{\mr{S},w}  &\simeq 
\frac{S_{w}^2}{2(\Ez_j {+} \Ez_k)^2}
+\frac{T_{w}^2}{2(\Ez_j {-} \Ez_k)^2} 
+ S_{w} T_{w} \bigl(\frac{1}{\Ez_j^2}{+}\frac{1}{\Ez_k^2}\bigr), \\[1ex]
e^{(w)}_{\mr{P},w}&\simeq
\Bigl(\frac{\theta_w/2}{T_{w}-S_{w}}\Bigr)^{\!2} \left[
\frac{S_{w}^4}{2(\Ez_j+\Ez_k)^2}  +\frac{T_{w}^4}{2(\Ez_j - \Ez_k)^2}
\right. \\
&\quad\quad\quad \left. +\,  S_{w}^2 T_{w}^2 \bigl(\frac{1}{\Ez_j^2}+\frac{1}{\Ez_k^2}\bigr)
+\frac{S_{w}^3 T_{w}-S_{w} T_{w}^3}{\Ez_j \Ez_k}\right],
 \end{aligned}
\end{equation}
where we have determined the evolution time $\tau$ from \Eqref{eq:mqgatetime}. We note that these are also the coherent errors associated with a CPhase gate on $w$. 

Given the significantly improved gate error estimations in Eqs.~\!\!\eqref{eq:err-decompose} and Eqs.~\!\!\eqref{eq:err2dots}, to accurately implement a multi-qubit gate, it is sufficient to require the quantization energy to differ across adjacent quantum dots $\abs{\Ez_j {-}\Ez_k}>J_{jk}$. This is a much weaker condition than requiring all $\{E_n\}$ to be non-degenerate. For randomly distributed quantum dots, it can be shown that the minimal detuning between near-neighbors scales as 
$O(1/N)$, as opposed to the $O(e^{-N})$ scaling for the full array. Therefore, the growth of coherent gate errors is  manageable as the system scales up. 
Furthermore, it is also possible to directly engineer large energy differences across neighboring dots, e.g., by interleaving two spices of quantum dots with different ranges of quantization energies. Noticeably, a similar arrangement is discussed for transform qubits, where engineered detuning following quasi-periodic frequency distributions are found to be optimal~\cite{Berke2022Transmon}. 
We plot this particular checkerboard style arrangement In \Figref{fig:class}(b). 
Following such device deign, the nearest-neighbor detuning is separated by an energy gap $\Delta$ that bounds the coherent infidelity by  
$\InF \lesssim O(J^2/\Delta^2)$ for all DC multi-qubits gates on the array,  or square of energy spectral power if pulse-shaping is used.

\begin{figure}[htbp]
 \centering
 \includegraphics*[width=10cm]{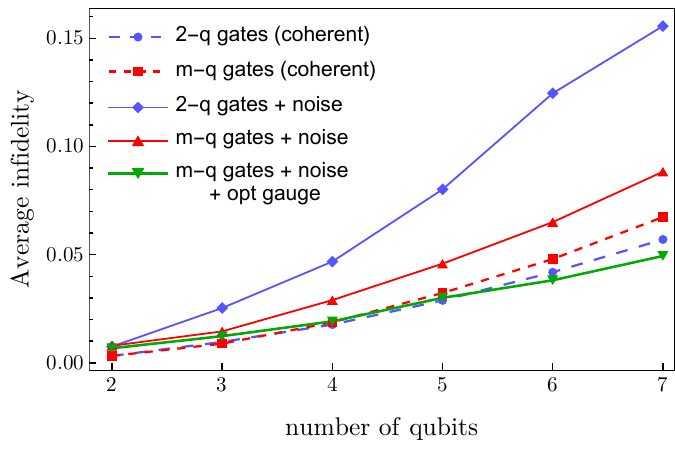}
 \caption{Benchmarking the $m$-qubit gate $\CZ_2\mr{Z}_3\cdots\mr{Z}_m$ with an equivalent circuit composed by $m$ applications of 2-qubit $\CZ$ gates. If only coherent errors are taken into account, multi-qubit gates suffer a slight performance hit compared to equivalent 2-qubit gates. If a small stochastic charge noise is considered, the multi-qubit gates will  outperform significant fidelity advantages over 2-qubit gates.  Under an optimal set of phase gauge determined in \Eqref{eq:optimal-corr}
The errors for multi-qubit gates can be further effectively suppressed.
}
 \label{fig:czzfd}
\end{figure}

A fair way to characteristic the noise level of a multi-qubit gate is to compare it with an equivalent circuit composed of  2-qubit gates. 
Equation \eqref{eq:err-decompose} establishes a set of inequality relations between these two parties, and suggests that a properly implemented multi-qubit DC gate will not be much more erroneous.  Meanwhile, due to shorter evolution time and unified control, a multi-qubit gate can typically have significantly smaller {incoherent} errors compared to the equivalent circuit of 2-qubit gates. This key advantage of multi-qubit gates is demonstrated in \Figref{fig:czzfd}, where we compare the average gate infidelity of  $m$-qubit MTCP gates $\CZ_2\mr{Z}_3\cdots\mr{Z}_m$ with the sequential products of 2-qubit $\CZ$ gates. The underlying array is of stellar topology with the Zeeman splitting $\Ez_1=1.2\ \mu$eV for the control qubits and 
$\Ez_j\in[0.3,0.4]\ \mu$eV for the target qubits $j\ge2$. The bonds are assumed to be identical with $J=0.05\ \mu$eV and $\tilde{s}/\tilde{t}=\tan(0.15\pi)$. 
If the interaction of system qubits with environment is ignored, namely the errors being completely coherent in nature, the infidelity of a MTCP gate is only slightly larger ($<120\%$)  than the products of 2-qubit gates. We model environmental charge noise by imposing stochastic fluctuations $\delta\Ez_j(t)\sim 0.01\mu$eV on the qubit quantization energies. This leads to dephasing noise on the qubits and is widely considered as a major source of error for spin qubits. Each point in \Figref{fig:czzfd} is obtained by averaging over 100 time steps for 1000 random ensembles. In such noisy case, it is found that MTCP gates have noticeably smaller overall errors than their 2-qubit-gate counterparts.  This fidelity advantage is increasing significant as the noise strength and/or the qubit number increases. 
Similar incoherent errors are often dominant in practical systems. Hence if possible, multi-qubit DC gates should be preferred over 2-qubit gates. 

\subsubsection{Optimal local gauge}
According to \Eqref{eq:inf-bound}, a considerable part of the coherent gate error can be attributed from the phase shifts  
$\{\tau \delta E^{(2+)}_n\}$ due to unattended higher-order energy corrections. Such undesired phase errors can accumulate in time and propagate through quantum circuits. Luckily, it is possible to suppress these excessive phase errors by applying \emph{additional} phase corrections to the phase gauge established for the ideal map. The notion of applying 
slightly different local phase corrections to enhance gate fidelity has already been studied for the CPhase gate \cite{Russ2018Highfidelity,Qi2024Spin}. 
In this subsection, we consider a similar fidelity optimization protocol, where one seek for an optimal set of local phase corrections to minimize the phase error  for a given multi-qubit DC gate. 

Previously, we have determined the phase gauge that transforms the ideal map $\Uideal$ into a useful quantum gate $G$ by the equality $G= Z(\bm\phi)\Uideal=\ee^{\ii \Phi}\Uideal$. 
According to \Eqref{eq:gauge-vec}, the gauge vector $\bm\Phi$ is related with the phase factors by a linear transformation,
\begin{equation}
\bm\Phi= \phi_\mr{g}+ K_{\!N} \bm\phi_\mr{loc},
\end{equation}
where $\bm\phi_\mr{loc}=(\phi_1,\phi_2,\cdots,\phi_N)^\tp$ is the vector of local phase factors determined from \Eqref{eq:cphase-gauge} and  $K_N$ is a $2^N\times N$ transformation matrix that depends only on $N$. It can be shown that the rows of $K_N$ are all the binary-digit vectors of length $N$. A simple proof of this property is given in Appendix~\ref{app:gaugevec}.  Hence $K_N$ can be efficiently calculated. 

With an alternative set of phase factors
$\bm\phi'$, the qubit frame map $\Uqb$ is transformed into the (imperfect) gate implementation 
$G'=Z(\bm\phi')\Uqb$.
Under this new gauge, the gate fidelity becomes
\begin{equation}
  \Fid(G,G')=\Fid(\Uideal, \ee^{\ii K_{\!N} \delta \bm\phi_\mr{loc}} \Uqb),
\end{equation}
where the additional local phase corrections are defined by $\delta\bm\phi_\mr{loc}=\bm\phi_\mr{loc}'-\bm\phi_\mr{loc}$. 
In such case, the gate infidelity upper bound is similar to \Eqref{eq:inf-bound}, except with the difference that the 
phase error is shifted to
\begin{equation}\label{eq:ePnew}
 e_\mr{P}'= \frac{1}{d}\, \norm{ \bm \zeta -K_{\!N} {\delta \bm\phi_\mr{loc}}}^2\equiv \frac{1}{d}\, \norm{ \bm \zeta'}^2,
\end{equation}
for the excessive phase vector defined by $\bm \zeta_n \equiv \tau \delta E_n^{(2+)}$. Therefore our goal can be formulated as determining an optimal vector 
$\delta \bm\phi_\mr{loc}$ such that the above $L_2$-norm is minimized. 
If $K_{\!N}^{-1}$ exists, it is easy to see that the phase error will vanish by taking ${\delta \bm\phi_\mr{loc}}=K_{\!N}^{-1} \bm \zeta$. But as an $2^{N}\times N$ matrix, $K_N$ cannot be inverted for $N\ge 2$. In such case, it follows from the theory of the Moore-Penrose pseudoinverse \cite{Ben-Israel1980Generalized} that the optimal choice of $\delta \bm\phi_\mr{loc}$ is given by,
\begin{equation}\label{eq:optimal-corr}
\begin{aligned}
 \delta\bm\phi_\mr{loc}^{(\mr{opt})}=K_{\!N}^+ \bm \zeta, 
\end{aligned}
\end{equation}
where $K_{\!N}^+\equiv(K_{\!N}^\tp K_{\!N})^{-1} K_{\!N}^\tp$
is the pseudoinverse matrix of $K_N$.  And the excessive phase vector becomes
$\bm \zeta' =(I-K_N K_N^+)\bm\zeta$ in the optimized gauge.

To demonstrate the effect of gauge optimization, let us consider the example of a triple-dot system with bonds $\bond{1,2}$, $\bond{2,3}$ and $\bond{3,1}$. 
Applying second-order perturbation theory, the excessive phase vector of the corresponding gate can be approximately calculated by
\begin{equation}\label{eq:exphase-3q}
 \bm \zeta \simeq \tau
\begin{pmatrix}
 \phantom{+}a_{12}+a_{23}+a_{31}+c\\
 \phantom{+}a_{12}+b_{23}-b_{31}-c\\
 \phantom{+}a_{31}+b_{12}-b_{23}-c\\
-a_{23}+b_{12}-b_{31}+c\\
 \phantom{+}a_{23}-b_{12}+b_{31}-c\\
-a_{31}-b_{12}+b_{23}+c\\
-a_{12}-b_{23}+b_{31}+c\\
-a_{12}-a_{23}-a_{31}-c
\end{pmatrix},
\end{equation}
where we introduce shorthands, 
\begin{equation}
\begin{aligned}
 &a_{ij} = \frac{S_{ij}^{2}}{\Ez_i+\Ez_j}+S_{ij} T_{ij}\Bigl(\frac{1}{\Ez_i}+\frac{1}{\Ez_j}\Bigr),\\
 &b_{ij} = \frac{T_{ij}^{2}}{\Ez_i-\Ez_j}+S_{ij} T_{ij}\Bigl(\frac{1}{\Ez_i}-\frac{1}{\Ez_j}\Bigr),\\
&c=\sum_{i=1}^3\frac{1}{4\Ez_i}\!\sum_{{j\neq k\neq i}}\! J_{ij} J_{ik}(\tilde{s}_{ij} \tilde{t}_{ij})
(\tilde{s}_{ik} \tilde{t}_{ik})^*.
\end{aligned}
\end{equation}
Applying optimal local phase corrections according to \Eqref{eq:optimal-corr}, the 
excessive phase vector becomes
$\bm \zeta' =(I-K_N K_N^+)\bm\zeta \simeq \tau c\times (1,-1,-1,1,-1,1,1,-1)^\tp$.
Compared with the uncorrected $\bm \zeta$ vector in \Eqref{eq:exphase-3q}, all $a_{ij}$ and $b_{ij}$ terms vanish. Since $\Ez_{i},\Ez_{j}\gg J_{ij}$, the $b_{ij}$ terms dominate over others terms. Hence the phase error is dramatically reduced in the optimal gauge. The effects of  gauge optimization  are also explicitly demonstrated in \Figref{fig:czzfd}, where we see a significant reduction in the gate infidelity if the addition phase corrections \Eqref{eq:optimal-corr} are applied to the noisy MTCP gates $\CZ_2\mr{Z}_2\cdots\mr{Z}_m$.

\section{Examples and applications}\label{sec:examples}
The rule of multi-qubit gate decomposition in \Eqref{eq:Uideal-decompose} allows us to conceive many multi-qubit DC gates convenient for quantum information processing tasks. 
Here we discuss some concrete examples and analyze their advantages over regular 2-qubit gates. Hopefully these discussions can inspire more practical applications of multi-qubit DC gates. 

\subsection{Three-qubit logical $Z$-gate}
Our first example is based on a triple-dot system. Instead of the regular chain topology, we consider the case where all three dots are all connected to each other as a ring, as shown in 
\Figref{fig:decompose}(b). 
Such kind of array topology and the corresponding gate have also been theoretically considered and experimentally tested in literatures \cite{Nguyen2025Singlestep,Acuna2024Coherenta}. 
From the gate decomposition theorem, the corresponding multi-qubit gate is equivalent to the product of three CPhase gates on each bond. Assuming that the bonds are homogeneous and that $\pi$-phase shifts are applied, the combined gate becomes 
\begin{equation}\label{eq:logical-Z}
\begin{split}
 G &= (\mr{C}_1 \mr{Z}_2) \cdot (\mr{C}_2 \mr{Z}_3) \cdot (\mr{C}_3 \mr{Z}_1)\\
&\repr \diag(1,1,1,-1,1,-1,-1,-1).
\end{split}
\end{equation} 
When acting the resulting gate on the computational states, 
the $\ket{000}$, $\ket{001}$, $\ket{010}$ and $\ket{100}$ states are unaffected while $\ket{011}$, $\ket{101}$, $\ket{110}$ and $\ket{111}$ experience $\pi$-phase (sign) flips. Namely, the gate $G$ is capable of distinguishing the majorly $\ket{0}$ states with the majorly $\ket{1}$ states. 
Meanwhile, the non-entangling gate $X_1 X_2 X_3$ correctly interchanges the majorly $\ket{0}$ states with the majorly $\ket{1}$ states and also anticommutes with $G$.
Based on this property, this trip-dot system can be considered as 
a hardware implementation for the repetition code for correcting bit-flip errors, with the logical $\ket{\bar 0}=\ket{000}$ and the logical $\ket{\bar 1}=\ket{111}$ states. The gate $G$ 
in \Eqref{eq:logical-Z} becomes the logical $\bar Z$ gate whereas 
 $X_1 X_2 X_3$ becomes the logical $\bar X$.
 We may compare $G$ with the simple product gate of $Z_1 Z_2 Z_3$, which also flips the sign of the  logical $\ket{\bar 1}$ state and anticommutes with $X_1X_2X_3$. But the $Z_1 Z_2 Z_3$ gate does not respect majority voting like $G$, and cannot serve as the logical $\bar Z$ as a result. Compare with an equivalent circuit using 3 applications of $\CZ$ gates, the multi-qubit gate $G$ require only a single-shot application and is therefore less error-prone. 
In comparison to the bitflip code presented here, we note that a similar phase-flip code for spin-qubits can be carried out by combing controlled-Z and controlled-S$^{-1}$ gate into a three-qubit Toffoli gate \cite{vanRiggelen-Doelman2024Coherent}.

\subsection{Simultaneous parity checks}
\begin{figure}[htbp]
 \centering
 \includegraphics[width=11cm]{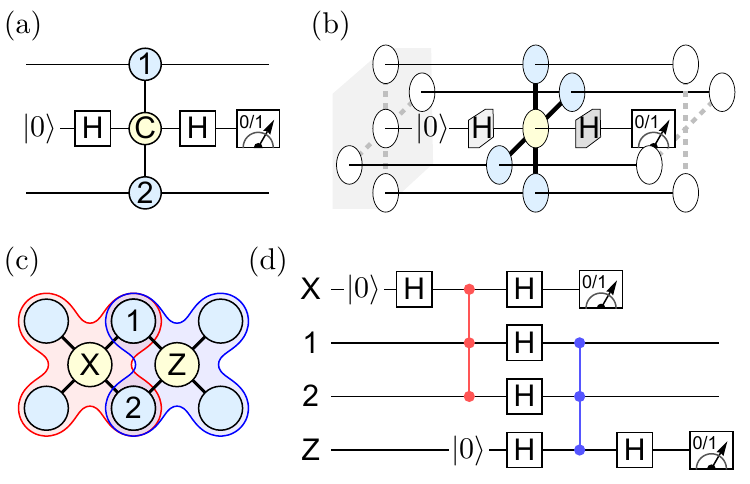}
 \caption{Example demonstrations of simultaneous parity checks through MTCP gates. (a) The quantum circuit  to perform parity check of the $Z_1Z_2$ operator. The circuit can also be slightly modified to perform parity check of the $X_1X_2$ operator.
(b) A two-dimension  parity check circuit that simultaneously check the parity of the four neighboring qubits by application of the $\CZ_1Z_2Z_3Z_4$ gate and measurement of the controlled qubit.  (c) A basic unit cell of the surface code that involves data qubits with $X$ and $Z$ parity check maps. Both of these maps can be efficiently carried out using MTCP gates. (d) The circuit diagram of the surface code stabilizing cycle for the shared data qubit 1 and 2. The red and blue vertical lines joining three circuit wires are an application of $\CZ_1Z_2$ gates, with the control qubits $X$ and $Z$ as  in (c).
}
 \label{fig:pcheck}
\end{figure}

The  ability to perform parity measurements on different qubits is a fundamental requirement for many quantum error correction codes \cite{Nielsen2010Quantum}. 
It turns out that the MTCP gates are particularly suitable for such parity measurement tasks. We demonstrate some explicit examples of MTCP as parity checkers in \Figref{fig:pcheck}. 

Let us first consider the circuit in  \Figref{fig:pcheck}(a) for a three-qubit system. This circuit can be broken down into three steps: (1) preparing the  middle qubit in the $\ket{+}$ state,
(2) applying a $\CZ_1Z_2$ gate with the middle qubit as control, and
(3) measuring the control qubit in the $\ket{\pm}$ basis.  It can be directly worked out that the measurement outcome will project the other two qubits into $\pm 1$ eigen-spaces of the $Z_1Z_2$ operator. Hence this circuit is tantamount to a simultaneous parity check of the upper and lower qubits. 
This circuit can also be adapted to measure the parity of the $X_1 X_2$ operator by appending extra Hadamard gates $H_1H_2$ after the $\CZ_1Z_2$ gate. 
In \Figref{fig:pcheck}(b), we extend the tree-qubit parity measurement circuit to a two-dimensional array. The measurement qubit now serves as the control qubit in the $\CZ_1Z_2Z_3Z_4$ gate. One can verify that the circuit measures the parity of the joint $Z$- or $X$- operator of the surrounding target qubits.
An equivalent parity check circuit can be certainly build using 2-qubit gates. Compared with the equivalent circuit with four CNOT gates  applied at different time step \cite{fowler2012surface}, the parity check circuits in \Figref{fig:pcheck}(b) only require one multi-qubit gate and the parity of the target qubits are measured simultaneously. 

The advantages of MTCP gates as simultaneous parity checks are obvious for large scale codes with many parity checks. In particular, the MTCP   parity checks can be used as an basic element for the constructing surface code. Following an example in Ref. \cite{fowler2012surface}, we consider a ``unit cell'' of surface code with one $X$ and one $Z$ parity check operators, as illustrated in \Figref{fig:pcheck}(c). We implement the $X$ and $Z$ parity checks with the red and blue colored MTCP gates respectively. These two gates are separated by single-qubit Hadamard gates for transforming the $\ket{0/1}$ basis with the $\ket{\pm}$ basis.  \Figref{fig:pcheck}(d) is a detailed breakdown of such simultaneous parity check for qubits 1 and 2, where the red and blue vertical lines are corresponding MTCP gates.  Following this circuit, one can verify that the final state for qubits 1 and 2 is stabilized to a simultaneous eigenstate of the $Z_1Z_2$ and $X_1X_2$ operator. Compared with equivalent circuit using 2-qubit gates, the MTCP gate approach requires significantly less entangling gates, and is capable of measuring the parity of all qubits simultaneously. This could dramatically reduce the  error rate associated with the parity measurement process.

\subsection{Fast array reversal}
Assuming that a quantum state is stored with an array of $N$ qubits, the task of array reversal is to ``flip'' the entire array such that each basis state  $\ket{\bm n}$ is swapped with its reversed state 
$\ket{\overleftarrow{\bm n}}$,
\begin{equation}\label{eq:arrreversal}
 \ket{\bm n}=\ket{n_1,n_2,\cdots,n_N}  \leftrightarrow \ket{n_N,\cdots,n_2,n_1}\equiv \ket{\overleftarrow{\bm n}},
\end{equation}
 e.g., $\ket{0111}\leftrightarrow \ket{1110}$. 
The map should be linear such that any quantum entanglement is preserved.
As a notable example, this array reversal step appears in the quantum Fourier transform circuit required for the Shor's algorithm \cite{Nielsen2010Quantum}.

The quantum circuit for reversing an arbitrary state typically breaks down into multiple application of 2-qubit swap gates between nearest-neighbors. Flipping a linear array of $N$ qubits requires $N(N-1)/2$ swap operations, with each swap typically made up of three CNOT gates \cite{Nielsen2010Quantum}. Although some swaps can be simultaneously performed to reduce wait time, the task still requires $O(N^2)$ steps to accomplish with 2-qubit gates.
However, such array reversal task can be achieved with only $N+1$ applications of a multi-qubit gate $L$, which can be decomposed as
\begin{equation}
 L = (\mr{C}_1 \mr{Z}_2) \cdot (\mr{C}_2 \mr{Z}_3) \cdots (\mr{C}_{N-1} \mr{Z}_{N}),
\end{equation}
hence it is directly implementable for spin qubits according to \Eqref{eq:Uideal-decompose}. 
Consider the following gate sequence that interleaves the $L$ gate with direct products of single-qubit Hadamard gates $H=H_1\otimes H_2\otimes\cdots \otimes H_N$,
\begin{equation}
 R = (H L) (H  L) \cdots (H L)= (HL)^{N+1}.
\end{equation}
Using the stabilizer group formalism \cite{Gottesman1997Stabilizer}, we can show that $R$ induces the array reversal transformation in \Eqref{eq:arrreversal}.
As a result, application of the multi-qubit gate accelerate the order reversal task to $O(N)$ time steps,  compared with the equivalent circuit of 2-qubit gates taking $O(N^2)$ steps.

\section{Summary and Outlooks}\label{sec:conclusion}
This article explores the set of multi-qubit DC gates naturally implementable on spin qubit arrays, covering key theoretical aspects of gate dynamics, fidelity estimation and optimizations, as well as advantageous applications.  The effective computational Hamiltonian is derived for spin qubit chips hosted by quantum dot arrays. Under time reversal symmetry, the spin-dependent tunneling coefficients for coupled pairs of quantum dots can be used to define entangled states, which induces anisotropic exchange coupling for spin qubits.  By perturbative expansion of the qubit-frame time evolution operator, we recognize the first-order dynamics as the ideal gates and higher order terms as coherent errors. 
When combined with local and global phase freedoms, these ideal gates  define the multi-qubit DC gate class. It is revealed that all multi-qubit DC gates are equivalent to simultaneous products of 2-qubit controlled phase gates up to a set of phase gauge transforms. 
On the other hand, using leading order perturbation, we find the coherent gate errors can be bounded and further suppressed in an optimal gauge.
Finally, we discuss some examples of multi-qubit gates and their applications. These examples showcase the advantages of our proposed multi-qubit gates over regular 2-qubit gates in speeding up quantum error correction and computational tasks. 

Our paper has uncovered an intriguing family of multi-qubit DC gates for spin qubits. A few important questions relating to their applications in practical systems can be investigated in future researches. For example, in multi-qubit arrays, interdot coupling strengths can be inhomogeneous and interdot crosstalk errors can become troubling. It is necessary to devise clever schemes to efficiently overcome this problem in an extended array. 
Next, as adiabatic pulse-shaping schemes have been analyzed for two-qubit gates, it is thus a natural question that whether and how pulse shaping can be transplant to multi-qubit gates. Additionally, since chiral effect can persist in certain arrays with a looped topology, it will be interesting to further investigate how such a effect can be used for novel quantum gates. Last but not least, based on the discussions of MTCP gates, it is suggested that the ability to accurately control the ratio between the spin-flipping and spin-conserving tunneling strength will be beneficial. It is left for further investigations on how this property can be utilized to devise efficient quantum control methods.

\section*{Acknowledgements}
J. Q. acknowledges support from Beijing Postdoctoral Research Foundation with
Grant No.~\!2023-zz-050 and the National Natural Science Foundation of China with Grant No.~12404562. 
H. Q. X. acknowledges support from  the National Natural Science Foundation of China (Grant Nos.~92165208 and 11874071).

\bibliographystyle{unsrt}
\bibliography{MQGates}

\newpage
\appendix

\section{Tunneling coefficients and weak time-reversal symmetry}\label{app:tunnel}
To establish the relations among tunneling coefficients, we 
utilize a time-reversal symmetry property of the tunneling Hamiltonian.

Let us recall that the time-reversal operator is define by $\widehat{T}=-\ii \sigma_y \widehat K$, with the Pauli matrix $\sigma_y$ acting on the spin part and the complex conjugation $\widehat K$ on the orbital part of a wave function. The action of time reversal on the field operators can be summarized by \cite{Sakurai2020Modern}
\begin{equation}
  \begin{aligned}
\widehat{T} a_{j\uparrow} \widehat{T}^\dagger &=  a_{j\downarrow},&
\widehat{T} a^\dagger_{j\uparrow} \widehat{T}^\dagger &=  a^\dagger_{j\downarrow},\\
\widehat{T} a_{j\downarrow} \widehat{T}^\dagger &=  -a_{j\uparrow}, &
\widehat{T} a^\dagger_{j\downarrow} \widehat{T}^\dagger &=  -a^\dagger_{j\uparrow}.
  \end{aligned}
\end{equation}
Examining the second-quantized Hamiltonian \Eqref{eq:FH-Hamiltonian}, we see that $H_{\mr{dot}}$ directly violate the time-reversal symmetry due to the sign reversal before the Zeeman energy,
\begin{equation}
  \widehat{T} H_{\mr{dot}} \widehat{T}^\dagger = \sum_{j}\sum_\sigma \left[ (\mu_j - {\sig(\sigma_j)}\,\frac{1}{2}\varepsilon_{j}) \, n_{j\sigma} + \frac{1}{2} U_j\, n_{j\sigma} n_{j\bar\sigma} \right]  \neq  H_{\mr{dot}},
\end{equation}
This is expected as an \emph{external} magnetic field explicitly breaks the time-reversal symmetry.
On the other hand, the tunneling Hamiltonian is \emph{internal} to the system and should be invariant under time reversal. In particular, at zero magnetic field where the system is fully closed, we strictly have $\widehat{T} H_{\mr{tun}}(\mb B=0) \widehat{T}^\dagger = H_{\mr{tun}}(\mb B=0) $. Furthermore, provide that the tunneling coefficients are independent of the external magnetic field strength, $H_{\mr{tun}}$ is unaffected by $\mb B$, then we have the full time-reversal symmetry condition
\begin{equation}\label{eq:trsym}
  H_{\mr{tun}}= \widehat{T}H_{\mr{tun}} \widehat{T}^\dagger,
\end{equation}
despite external magnetic field. 

Recall that the tunneling coefficients are defined during second-quantization of the single-body Hamiltonian of the array $H_\mr{arr}$ by the matrix elements,
$t^{jk}_{\sigma\sigma'}\equiv\matele{\Phi_{j\sigma}}{H_\mr{arr}}{\Phi_{k\sigma'}}$, 
where $\{\Phi_{j\sigma}\}$ is the low energy basis for the array used to define \Eqref{eq:FH-Hamiltonian}.  
By construction, we can split $H_\mr{arr}$ into a local Hamiltonian 
$H_{\mr{loc},j}$ that only contains the harmonic potential profile around dot $j$, 
in addition to the potential difference $\Delta V_{j}$ between the local potential  and the full array, 
$H_\mr{arr} = H_{\mr{loc},j}+\Delta V_{j} = H_{\mr{loc},k}+\Delta V_{k}
  \equiv \frac{1}{2} ( H_{\mr{loc},j}+ H_{\mr{loc},k}) + \Delta V_{jk}$. Then the tunneling coefficients can be approximated by \cite{Qi2024Spin},
\begin{equation}\label{eq:tjk}
  \begin{aligned}
    t^{jk}_{\sigma\sigma'}\simeq \braket{\phi_{j\sigma}}{\phi_{k\sigma'}} 
\left(\frac{1}{2}\mu_{k} -\frac{1}{2}\mu_{j} +v_{jk}\right),
\end{aligned}
\end{equation}
where $\ket{\phi_{j\sigma}}=\ee^{\ii \mb{p}\cdot\mb{r}_j} \ket{0}\ket{\sigma}$ is the local eigenstate for $H_{\mr{loc},j}$ and $v_{jk}=\matele{0}{\Delta V_{jk}}{0}$ characterizes the potential barrier energy between the dots. Such approximation is valid provided that the Zeeman energies $\Ez_j$ and $\Ez_k$ are much smaller than the orbital potential barrier energy $\abs{v_{jk}}$. 
Now that the local states $\ket{\phi_{j\sigma}}$ and $\ket{\phi_{k\sigma'}}$ are independent of $\abs{\mb B}$ up to the first order perturbation.
Neither is the interdot barrier energy dependent on $\abs{\mb B}$. Hence we find the partial derivative 
\begin{equation}\label{eq:tpartialB}
  \frac{\partial}{\partial \abs{\mb B}} \, t^{jk}_{\sigma'} 
= O\left(({\Ez}/{\abs{v}})^2\right) \ll 1,
\end{equation}
Hence in the working regime of a typical spin qubit device, the tunneling coefficients can be regarded as  being independent of the magnetic field. This justifies the time-reversal symmetry condition. 

Applying the time-reversal symmetry condition [\Eqref{eq:trsym}], we can equate 
\begin{equation}
t^{jk}_{\uparrow\uparrow} = (t^{jk}_{\downarrow\downarrow})^{*},\quad
t^{kj}_{\uparrow\uparrow} = (t^{kj}_{\downarrow\downarrow})^{*},\quad 
t^{jk}_{\uparrow\downarrow} = - ( t^{jk}_{\downarrow\uparrow})^*,\quad 
t^{kj}_{\uparrow\downarrow} = - ( t^{kj}_{\downarrow\uparrow})^*.
\end{equation}
Another set of relation follows from the Hermicity $H_{\mr{tun}}^\dagger = H_{\mr{tun}}$,
\begin{equation}
t^{jk}_{\uparrow\uparrow} = (t^{kj}_{\uparrow\uparrow})^{*},\quad 
t^{jk}_{\downarrow\downarrow} = (t^{kj}_{\downarrow\downarrow})^{*},\quad
t^{jk}_{\uparrow\downarrow} = (t^{kj}_{\downarrow\uparrow})^{*},\quad 
t^{jk}_{\downarrow\uparrow} = (t^{kj}_{\uparrow\downarrow})^{*}.
\end{equation}
Combining these two sets of relations, it follows that there are only two independent tunneling coefficients, representing the spin-conserved and spin-flipped processes respectively
\begin{equation}
  \begin{aligned}
    t^{jk}_{\uparrow\uparrow} &= (t^{jk}_{\downarrow\downarrow})^{*} 
    = t^{kj}_{\downarrow\downarrow} =( t^{kj}_{\uparrow\uparrow})^* \\ 
    t^{jk}_{\uparrow\downarrow} &= -(t^{jk}_{\downarrow\uparrow})^{*} 
    = -t^{kj}_{\uparrow\downarrow} =( t^{kj}_{\downarrow\uparrow})^* .
  \end{aligned}
\end{equation} 
This produces \Eqref{eq:tun-coeffs} in the main text.

\section{Computational Hamiltonian}\label{app:compH}
To describe the dynamics with multiple spin qubits, we must derive 
a matrix representation of the low-energy Hamiltonian \Eqref{eq:FH-Hamiltonian} using multi-body basis states. These states are anti-symmetrized product states of the single-body wave functions.
For example, a system of three quantum dots all in the spin-up state is 
specified by the antisymmetric wave function:
\begin{equation}\label{eq:state_le}
  \ket{\uparrow\uparrow\uparrow} 
=\widehat{A}\,\Bigl(
\ket{\Phi_{1\uparrow}}\ket{\Phi_{2\uparrow}} \ket{\Phi_{3\uparrow}} \Bigr) 
=\frac{1}{\sqrt{3!}}
  \begin{vmatrix} 
    \ket{\Phi_{1\uparrow}}_1 & \ket{\Phi_{2\uparrow}}_1 & \ket{\Phi_{3\uparrow}}_1\\
    \ket{\Phi_{1\uparrow}}_2 & \ket{\Phi_{2\uparrow}}_2 & \ket{\Phi_{3\uparrow}}_2\\
    \ket{\Phi_{1\uparrow}}_3 & \ket{\Phi_{2\uparrow}}_3 & \ket{\Phi_{3\uparrow}}_3
  \end{vmatrix}
,
\end{equation}
where $\widehat{A}$ denotes the antisymmetrization operator; the subscripts for kets in the Slater determinant explicitly label the charge carriers. But such notation is irrelevant after antisymmetrization.
By considering all combinations of single body wave functions, we also allow two charge carriers occupying the same dot. Restring to the ground orbital states, the only possibilities are that of antiparallel states within a dot. We denote such state using the letter `S' in suggestion of a singlet state. But the actual wave function differ from a plain singlet as the antisymmetrization is performed over all fermions instead of just two. For example, we define the shorthand 
\begin{equation}\label{eq:state_he}
  \ket{0\,\mr{S}\!\uparrow} 
= \widehat{A}\,\Bigl(
\ket{\Phi_{2\uparrow}}\ket{\Phi_{2\downarrow}} \ket{\Phi_{3\uparrow}} \Bigr) 
 =\frac{1}{\sqrt{3!}}
  \begin{vmatrix} 
    \ket{\Phi_{2\uparrow}}_1 & \ket{\Phi_{2\downarrow}}_1 & \ket{\Phi_{3\uparrow}}_1\\
    \ket{\Phi_{2\uparrow}}_2 & \ket{\Phi_{2\downarrow}}_2 & \ket{\Phi_{3\uparrow}}_2\\
    \ket{\Phi_{2\uparrow}}_3 & \ket{\Phi_{2\downarrow}}_3 & \ket{\Phi_{3\uparrow}}_3
  \end{vmatrix},
\end{equation}
where the label '0' indicates that the first dot is unoccupied.
In the followings, we will assume all such multi-body states are defined in such antisymmetric manner. 

The states in \Eqref{eq:state_le} and \Eqref{eq:state_he} are examples of the half-filling states and doubly-occupied states. Splitting the multi-body wave function basis according to  the half-filling states and the doubly-occupied states, the Hamiltonian carries the following representation in the combined Hilbert space 
\begin{equation}\label{eq:H-block}
 H = 
\left[
\begin{array}{c c c|c c c}
\ddots&&&&&\\
 & H_{\mathrm{low}} & & & T^\dagger &\\ 
&&\ddots&&&\\\hline
&&&\ddots&& \\
& T & & & H_{\mathrm{high}} & \\
&&&&& \ddots
 \end{array}
\right]
\begin{array}{l}
\left.\lefteqn{\phantom{\begin{array}{c}\ddots\\H_x\\\ddots\end{array}}}\right\}
\text{half-filling states}\\
\left.\lefteqn{\phantom{\begin{array}{c}\ddots\\H_x\\\ddots\end{array}}}\right\}
\text{doubly-occupied states}
\end{array},
\end{equation}
where the diagonal blocks $H_\mathrm{low}$ and  $H_\mathrm{high}$ arise from the dot Hamiltonian $H_\mr{dot}$ while the anti-diagonal block $T$ results from the interdot tunneling $H_\mr{tun}$.

Since all $\ket{\Phi_{j\sigma}}$ are just eigenstates of $H_\mr{dot}$
and that the antisymmetrization operator commute with the second  quantized Hamiltonian, $H_\mr{dot}$ contributes to the diagonal matrix elements. For the half-filling states, we have
\begin{equation}\label{eq:Hd-lstates}
\begin{aligned}
H_\mr{dot} &\, \ket{\sigma_1 \sigma_2 \cdots \sigma_N} = \widehat{A} H_\mr{dot} \, \ket{\Phi_{1,\sigma_1}} \ket{\Phi_{2,\sigma_2}}  \cdots \ket{\Phi_{N,\sigma_N}} 
= \sum_j \varepsilon_{j\sigma_j} \ket{\sigma_1 \sigma_2 \cdots \sigma_N}\\
&=\Bigl(\sum_{j}\mu_j + \sum_{j} \frac{1}{2}
{\sig(\sigma_j)}\varepsilon_{\mr{Z},j}\Bigr)  \ket{\sigma_1 \sigma_2 \cdots \sigma_N} ,
\end{aligned}
\end{equation}
On the other hand, when acting $H_\mr{dot}$ on the doubly-occupied states, the  Coulomb charging energy $U_j$ appears on the diagonal elements, 
\begin{equation}\label{eq:Hd-hstates}
 H_\mr{dot}\, \ket{\cdots 0_n \cdots \mr{S}_m \cdots}\simeq \Bigl( \sum_j \mu_j +(\mu_m-\mu_n)+ U_m  \Bigr) \ket{\cdots 0_n \cdots \mr{S}_m \cdots},
\end{equation}
where we have neglected the Zeeman energy contributions in \Eqref{eq:Hd-hstates}.
Typically the Coulomb charging energy is much larger than the Zeeman energy on each dots, therefore the half-filling states and doubly-occupied states divide the Hilbert space into low energy and high energy subspaces, as suggests by the Hamiltonian subscripts. 
We can take out the common energy shift due to the summation of all chemical potential terms $\sum_{j}\mu_j$ in \Eqref{eq:Hd-lstates} and \Eqref{eq:Hd-hstates}.
For the low-energy subspace in particular, we can identify 
\begin{equation}\label{eq:Hlow}
 H_\mr{low} = \sum_{j} \frac{1}{2}\Ez_{j} \sigma^\mr{Z}_{j},
\end{equation}
This defines the qubit Hamiltonian for the quantum dot array.

The tunneling Hamiltonian $H_\mr{tun}$  contributes to the off-diagonal matrix elements in \Eqref{eq:H-block}.  
Special attention must be paid to take account of the the antisymmetrization of wave functions. 
For example, let us consider a triple-dot system, with dot 1,2,3. The dots are arranged linearly such that direct tunneling is only among dot 1,2 and dot 2,3. The tunneling Hamiltonian in this case is
\begin{equation}
\begin{aligned}
H_\mr{tun} 
&= t_{1}\, a^+_{1\uparrow}a_{2\uparrow}
+t^*_{1}\, a^+_{1\downarrow}a_{2\downarrow} 
+s_{1}\, a^+_{1\uparrow}a_{2\downarrow}
-s^*_{1}\, a^+_{1\downarrow}a_{2\uparrow} 
+h.c. \\
&+ t_{2}\, a^+_{2\uparrow}a_{3\uparrow}
+t^*_{2}\, a^+_{2\downarrow}a_{3\downarrow} 
+s_{2}\, a^+_{2\uparrow}a_{3\downarrow}
-s^*_{2}\, a^+_{2\downarrow}a_{3\uparrow} 
+h.c.
\end{aligned}
\end{equation}
where $t_{1}$ and $s_1$ are the tunneling coefficients between dot 1 and 2, $t_2$ and $s_2$ are the tunneling coefficients between dot 2 and 3.
We can easily work out its action with the help of the antisymmetrization operator. For example,
\begin{align}
H_\mr{tun}&\, \ket{\uparrow\uparrow\uparrow}
= H_\mr{tun} \,\widehat{A}\,\ket{\Phi_{1\uparrow}}\ket{\Phi_{2\uparrow}} \ket{\Phi_{3\uparrow}}
=  \widehat{A}\, H_\mr{tun} \,\ket{\Phi_{1\uparrow}}\ket{\Phi_{2\uparrow}} \ket{\Phi_{3\uparrow}} \notag \\
& =   \widehat{A}\, \left(-s^*_{1}\, a^+_{1\downarrow}a_{2\uparrow}  
+ s^*_{1}\, a^+_{2\downarrow}a_{1\uparrow}  
-s^*_{2}\, a^+_{2\downarrow}a_{3\uparrow} 
+s_2^*\, a^+_{3\downarrow}a_{2\uparrow} \right)\ket{\Phi_{1\uparrow}}\ket{\Phi_{2\uparrow}} \ket{\Phi_{3\uparrow}} \notag  \\
&=   \widehat{A}\, \left(
-s^*_{1}\,\ket{\Phi_{1\uparrow}}\ket{\Phi_{1\downarrow}} \ket{\Phi_{3\uparrow}} 
+ s^*_{1}\,\ket{\Phi_{2\downarrow}}\ket{\Phi_{2\uparrow}} \ket{\Phi_{3\uparrow}}  
-s^*_{2}\,\ket{\Phi_{1\uparrow}}\ket{\Phi_{2\uparrow}} \ket{\Phi_{2\downarrow}} 
+s_2^*\,\ket{\Phi_{1\uparrow}}\ket{\Phi_{3\downarrow}} \ket{\Phi_{3\uparrow}}  \right) \notag \\
  &= -s_1^* \ket{\mr{S}\,0\!\uparrow}-s_1^*\ket{0\,\mr{S}\!\uparrow}
  -s_2^* \ket{\uparrow\! 0\,\mr{S}}-s_2^*\ket{\uparrow\!\mr{S}\,0}.
\end{align}
Notice the sign flip for the second and forth term due to application of  the antisymmetrization operator. Carrying out similar calculations, we can explicitly obtain the off-diagonal block for the triple-dot chain, 
\begin{equation}\label{eq:Tmat}
   T = \begin{array} {cc}
\begin{array}{cccccccc} \uparrow\uparrow\uparrow & \uparrow\uparrow\downarrow &
\uparrow\downarrow\uparrow & \uparrow\downarrow\downarrow &
\downarrow\uparrow\uparrow & \downarrow\uparrow\downarrow &
\downarrow\downarrow\uparrow & \downarrow\downarrow\downarrow 
\end{array} & \\
\left[ \begin{array}{cccccccc} 
-s_1^*&&t_1^* & & -t_1&&-s_1&\\
&-s_1^*& &t_1^* & &-t_1&&-s_1 \\
-s_1^*&&t_1^* & & -t_1&&-s_1&\\
&-s_1^*& &t_1^* & &-t_1&&-s_1\\
-s_2^*&t_2^*&-t_2 &-s_2 & &&&\\
-s_2^*&t_2^*&-t_2 &-s_2 & &&&\\
&& & & -s_2^*&t_2^*&-t_2 &-s_2 \\
&& & & -s_2^*&t_2^*&-t_2 &-s_2 
\end{array} \right] & \hspace{-1em} \begin{array}{c}
\mr{S}\,0\!\uparrow\\
\mr{S}\,0\!\uparrow \\
0\,\mr{S} \!\uparrow\\
0\,\mr{S} \!\uparrow\\
\uparrow\!\mr{S}\,0\\
\uparrow\!\mr{S}\,0\\
\downarrow\!0\,\mr{S} \\
\downarrow\!0\,\mr{S}
\end{array}
\end{array}
\end{equation}

To incorporate the exchange interaction between spins on different sites, we must properly account for the virtual process where low-energy states briefly tunnels to the high-energy states and back. 
The net effects on the low-energy subspace can be derived using the Schrieffer-Wolff transformation.  
The idea is to apply a basis change $\ee^S$ to bring the Hamiltonian in \Eqref{eq:H-block} into block-diagonal form, and the computational Hamiltonian is defined by 
$H_\mr{comp} = \mathcal{P} \ee^{S} H \ee^{-S}$, where $\mathcal{P}$ is the projection operator on the half-filling subspace.
Assuming $H=H_0+V$, where 
\begin{equation}
 H_0 = \begin{bmatrix}
   H_\mr{low} & 0 \\
0   & H_\mr{high} 
 \end{bmatrix},\quad 
V = \begin{bmatrix}
   0 & T^\dagger \\
T   & 0
 \end{bmatrix}
\end{equation}
 represents the diagonal and off-diagonal blocks of the full $H$ in \Eqref{eq:H-block}. An matrix $S$ that satisfies $[H_0,S]=V$ can  transform the Hamiltonian into block-diagonal form up to the fourth order,  giving 
\begin{equation}
 H_\mr{comp} = H_\mr{low} + \mathcal{P} \frac{1}{2}[S,V] + O((t/U)^4),
\end{equation} 
where $t$ and $U$ stands for the characteristic tunneling energy and on-site Coulomb charging energy. The ratio between the two is assumed very small such that the first two terms give an accurate depiction of the computation Hamiltonian. As the first term is already derive in \Eqref{eq:Hlow}, the goal is to derive a suitable expression for the second term.  As the diagonal block $H_0$ is in fact fully diagonal, we can explicit construct $S$ by its elements 
\begin{equation}\label{eq:SW}
 S_{ii}=0, \quad S_{ij}=\frac{V_{ij}}{(H_0)_{ii}-(H_0)_{jj}} \quad (i\neq j).
\end{equation}
One can obtain out the resulting Hamiltonian by substituting in expressions for $H_0$ and $V$.

Let us consider the triple-dot chain example. The $V$ matrix is already specified by the the off-diagonal block in \Eqref{eq:Tmat}. Under the same ordering of basis states, we also have the low-energy diagonal block
\begin{multline}
 H_{\mr{low}}= \frac{1}{2} \diag\bigl(
\Ez_1 +  \Ez_2+  \Ez_3,\,
\Ez_1 +  \Ez_2-  \Ez_3,\,
\Ez_1 - \Ez_2+  \Ez_3,\,
\Ez_1 -  \Ez_2-  \Ez_3,\\
-\Ez_1 +  \Ez_2+  \Ez_3,\,
-\Ez_1 +  \Ez_2-  \Ez_3,\,
-\Ez_1 - \Ez_2+  \Ez_3,\,
-\Ez_1 -  \Ez_2-  \Ez_3\bigr),
\end{multline}
and the low-energy diagonal block
\begin{multline}
 H_{\mr{high}}=  \diag\bigl(
U_1+\mu_1- \mu_2 +\tfrac{1}{2}\Ez_3,\,
U_1+\mu_1- \mu_2 -\tfrac{1}{2}\Ez_3,\,
U_2-\mu_1+ \mu_2 +\tfrac{1}{2}\Ez_3,\,
U_2-\mu_1+ \mu_2 -\tfrac{1}{2}\Ez_3,\\
U_2+\mu_2- \mu_3 +\tfrac{1}{2}\Ez_1,\,
U_3-\mu_2+ \mu_3 +\tfrac{1}{2}\Ez_1,\,
U_2+\mu_2- \mu_3 -\tfrac{1}{2}\Ez_1,\,
U_3-\mu_2+ \mu_3 -\tfrac{1}{2}\Ez_1 \bigr).
\end{multline}
After some algebra, we can explicitly work out the commutator, which after projection onto the low-energy space can be decomposed as 
\begin{equation}
  \mathcal{P} \frac{1}{2}[S,V] = H_{\mr{ex,12}} \otimes I_3
+  I_1 \otimes H_{\mr{ex,23}},
\end{equation}
where   $H_{\mr{ex,12}}$ and $H_{\mr{ex,23}}$ are four-dimensional matrices on the subspace of the qubit-pair 1,2 and 2,3 respectively; $I_1$ and $I_3$ are identity matrices on qubit-1 and qubit-3 subspace.  The matrix elements of  $H_{\mr{ex,12}}$ is explicitly given by 
\begin{equation}
 (H_{\mr{ex,12}})_{ij} = -\frac{1}{2} (\bm j_{12,i} + \bm j_{12,j}) \, \bm \xi_{12,i}\bm \xi_{12,j}^* ,
\end{equation}
for the vector $\bm j_{12}$ defined by 
\begin{equation}\label{eq:j12}
\bm j_{12}=
 \begin{pmatrix}
 \frac{1}{U_1+\mu_1-\mu_2 -\Ez_1/2-\Ez_2/2} + \frac{1}{U_2+\mu_2-\mu_1 -\Ez_1/2-\Ez_2/2}\\
 \frac{1}{U_1+\mu_1-\mu_2 -\Ez_1/2+\Ez_2/2} + \frac{1}{U_2+\mu_2-\mu_1 -\Ez_1/2+\Ez_2/2}\\
 \frac{1}{U_1+\mu_1-\mu_2 +\Ez_1/2-\Ez_2/2} + \frac{1}{U_2+\mu_2-\mu_1 +\Ez_1/2-\Ez_2/2}\\
 \frac{1}{U_1+\mu_1-\mu_2 -\Ez_1/2-\Ez_2/2} + \frac{1}{U_2+\mu_2-\mu_1 +\Ez_1/2+\Ez_2/2},
\end{pmatrix}
\end{equation}
and the vector $\bm \xi_{12}\equiv (s_1,-t_1,t_1^*,s_1^*)$.
As the Zeeman energy $\Ez$ is small compared to both $U$ and $\mu$, the four components in \Eqref{eq:j12} are almost identical to each other. It is customary to introduce the exchange energy between dot 1,2 by 
\begin{equation}\label{eq:J12}
 J_{12} \equiv \frac{(\abs{s_1}^2+\abs{t_1}^2)}{2}  \left(\frac{1}{U_1+\mu_1-\mu_2} + \frac{1}{U_2+\mu_2-\mu_1}\right).
\end{equation}
Then we can simplify the exchange Hamiltonian into 
\begin{equation}
 H_{\mr{ex,12}} \simeq - J_{12} \ket{\xi_{12}}\bra{\xi_{12}},
\end{equation}
for the entangled state defined on the Hilbert space of dot 1,2 by 
\begin{equation}\label{eq:xi12}
 \begin{aligned}
  \ket{\xi_{12}} 
& \equiv \frac{1}{\sqrt{2}} \left(
 \tilde s_1 \ket{\uparrow\uparrow}_{12} -
 \tilde  t_1 \ket{\uparrow\uparrow}_{12} +  \tilde  t_1^* \ket{\uparrow\uparrow}_{12}  + \tilde   s_1^* \ket{\uparrow\uparrow}_{12} \right),
 \end{aligned}
\end{equation}
for the dimensionless and normalized tunneling coefficients $\tilde s$ and $\tilde{t}$ satisfying $\abs{\tilde s_1}^2+\abs{\tilde{t}_1}^2=1$. 
The other exchange Hamiltonian $H_{\mr{ex},23}$ is defined similarly, with the replacement of pair 1,2 with pair 2,3. This lead to the total exchange Hamiltonian 
\begin{equation}\label{eq:Hex-3}
 H_{\mr{ex}} =  - J_{12} \ket{\xi_{12}}\bra{\xi_{12}}  - J_{23} \ket{\xi_{23}}\bra{\xi_{23}}.
\end{equation}

This result for triple-dot chain extends an earlier result for double-dot system~\cite{Qi2024Spin}, and inspires us to formulate a more general form of the exchange Hamiltonian for an arbitrary array of quantum dots 
\begin{equation}\label{eq:Hex-general}
 H_{\mr{ex}} = - \sum_w J_w \ket{\xi_w}\bra{\xi_w},
\end{equation}
where the summation index $w$ ranges over all pairs of directly connected quantum dots. Both the exchange energy $J_w$ and the state $\ket{\xi_w}$ are defined similarly as in \Eqref{eq:J12} and \Eqref{eq:xi12}. That is, we attach entangled states for every connecting bonds of the array, and sum up all the exchange Hamiltonian independently. 

 We can prove this conjecture \eqref{eq:Hex-general} by examining the steps that leads to \eqref{eq:Hex-3}. The key observation is that the tunneling coefficients for different bonds correspond to different matrix elements in distinctive matrix blocks. Therefore we can split the $V$ and $S$ matrix by 
\begin{equation}
 V = \sum_w V_w,\quad  S = \sum_w S_w,
\end{equation}
where $V_w$ consists of only elements proportional to $s_w$ or $t_w$, 
and $S_w$ is defined by $V_w$ according to \Eqref{eq:SW}.
Specifically, we can represent  $V_w$ by 
\begin{equation}
 V_w = \sum_{n,j} v^w_{n,j} \ket{\Omega^{w}_{n,j}}\bra{n} + h.c.,
\end{equation} 
where $\ket{n}$ is from the set of half-filling states and $\ket{\Omega^{w}_{n,j}}$ is from the set of doubly-occupied states, $v^w_{n,j}$ is proportional to the tunneling coefficient  $s_w$ or $t_w$. As different tunneling process maps $\ket{n}$ to different doubly-occupied states, 
we have 
\begin{equation}
 \braket{\Omega_{n,j}^w}{\Omega_{m,k}^v}\propto \delta_{wv}.
\end{equation}
As $S_w$ is defined element-wise by $V_w$, we also have the representation 
\begin{equation}
 S_w = \sum_{n,j} s^w_{n,j} \ket{\Omega^{w}_{n,j}}\bra{n} + h.c.,
\end{equation} 
for some complex coefficient $s^w_{n,j}$. It is now straightforward to verify that 
\begin{equation}
 \mathcal{P} S_w V_v = \mathcal{P} V_v S_w = \mathcal{P} [S_w,V_v] =0 \quad \text{for } w\neq v.
\end{equation}
A consequence for this commutativity is that the full basis rotation for the Schrieffer-Wolff transformation is now decomposed into successive rotations responsible for the tunneling coefficients for each connecting bonds. As each pair of commutator
for $S_w$  and $V_w$ can be worked out in the subspace of the pair $w$, which is already solved for the double-dot case, we have 
\begin{equation}
 H_\mr{ex} = \sum_{w,v} \mathcal{P} \frac{1}{2} [S_w,V_v]
=  \sum_w \mathcal{P} \frac{1}{2} [S_w,V_w]
= -\sum_w  J_w \ket{\xi_w}\bra{\xi_w}. 
\end{equation}
\qed

\section{Axial symmetry of the exchange coupling tensor}\label{app:axisym}

We recall that a vector $\bm{a}$ is reflected by a 3-vector $\bm{n}$ by 
\begin{equation}
  \bm{a} \to R_n \bm{a} = (-\mb{I}+2 \bm{n}\bm{n}^T ) \cdot \bm{a} 
\end{equation}
An axially symmetry exchange interaction satisfies
$R_n \mathcal{J} R_n = \mathcal{J}$ 
for some $\bm{n}$.
Here $ \mathcal{J}$ is the interdot exchange tensor when the exchange Hamiltonian is represented as 
$
H_{\mr{ex}} = \mb{S_1} \mathcal{J} \mb{S_2} 
$
with each $\bm{S_i}=1/2 (\sigma_i^x,\sigma_i^y,\sigma_i^z)$.

In our notation, the exchange coupling for each bond is resented in the spin space by 
$
H_{\mr{ex}} = - J \ket{\xi}\bra{\xi}
$
with 
$$
\ket{\xi} = 1/\sqrt{2} (\tilde{s},-\tilde t, \tilde t^*, \tilde s^*)^T
$$
in the spin basis $\{\ket{\uparrow\uparrow},\ket{\uparrow\downarrow},\ket{\downarrow\uparrow},\ket{\downarrow\downarrow}\}$. The dimensionless coefficients are normalized by $\abs{\tilde t}^2+\abs{\tilde s}^2=1$.
Apparently, only the relative phase between $\tilde t$ and $\tilde s$ is relevant. Without loss of generality, we can parameterize $\tilde t$ and $\tilde s$ by 
$$
\tilde{t} = \cos(\gamma)+ \ii \sin(\gamma)\cos(\vartheta), \quad
\tilde{s} =  \ii  \sin(\gamma)\sin(\vartheta)
$$
for some real number $\gamma$ and $\vartheta$.

To convert this entanglement state representation to tensor representation, we use 
$$
\mathcal{J}_{\alpha\beta} = \operatorname{Tr}(\sigma^\alpha \otimes \sigma^\beta H_\mr{ex}), \quad \alpha,\beta =x,y,z.
$$
This produces 
$$
\mathcal{J} = J \left(
\begin{array}{ccc}
 \cos ^2(\gamma )-\sin ^2(\gamma ) \cos (2 \vartheta ) & \sin (2 \gamma ) \cos (\vartheta ) & \sin ^2(\gamma ) \sin (2 \vartheta ) \\
 -2 \sin (\gamma ) \cos (\gamma ) \cos (\vartheta ) & \cos (2 \gamma ) & \sin (2 \gamma ) \sin (\vartheta ) \\
 \sin ^2(\gamma ) \sin (2 \vartheta ) & -2 \sin (\gamma ) \cos (\gamma ) \sin (\vartheta ) & \sin ^2(\gamma ) \cos (2 \vartheta )+\cos ^2(\gamma ) \\
\end{array}
\right)
$$
We now define a unit vector and its reflection operator
$$
\bm{n} = (\sin (\vartheta ),0,\cos (\vartheta ) )^T, \quad 
R_n= (-\mb{I}+2 \bm{n}\bm{n}^T )
$$
One can explicitly verify that $R_n \mathcal{J} R_n^{-1} = \mathcal{J}$ after some straightforward algebra.

\section{Reflective symmetry of the array vector}\label{app:refsym}
To reverse the elemental order of a vector, one can apply the $X$ matrix which is defined by 1's on the anti-diagonal elements and $0$'s elsewhere. That is 
\begin{equation}
  X \bm a = \overleftarrow{\bm a}.
\end{equation}
Applying the $X$ operator to the array vector, we find 
\begin{equation}
\begin{aligned}
    X \bm \Lambda &= X \bigotimes_w \bm\Lambda_w 
=  \sum_w X (\bm\Lambda_w \otimes \bm{1}_{w\perp})\\
&= \sum_w X_w \bm\Lambda_w  \otimes X_{w\perp} \bm{1}_{w\perp}
=  \sum_w \bm \Lambda_w  \otimes \bm{1}_{w\perp} =\bm \Lambda,
\end{aligned}
\end{equation}
where $\bm{1}_{w\perp}$ is a vector of $1$'s defined on the orthogonal space of bond $w$ and we have used the reflective symmetry of the bond vectors $X_w \bm\Lambda_w=\bm\Lambda_w$. 
Hence we prove that the array vector $\mb\Lambda$ is also reflectively symmetric.

\section{Calculating the gauge transformation matrix}\label{app:gaugevec}
According to \Eqref{eq:gauge-vec},
The gauge transformation matrix is defined by the relation
\begin{equation}
\bigoplus_{j=1}^N (0,\phi_j) \equiv K_{\!N} \bm\phi_\mr{loc}.
\end{equation}
Here $\bm\phi_\mr{loc}=(\phi_1,\phi_2,\cdots,\phi_N)^\tp$ is the local phase vector.
Apparently, we have
\begin{equation}
K_1 = \begin{pmatrix}
  0 \\
1
\end{pmatrix},
\quad
K_2=\begin{pmatrix}
0 & 0 \\
0 & 1\\
1 & 0 \\
1 & 1
\end{pmatrix}.
\end{equation}
For $N\ge 2$, we can expand the Kronecker sum over the first dot
\begin{equation}
K_{\!N} \bm\phi_\mr{loc} =(0,\phi_1) +\bigoplus_{j=2}^N (0,\phi_j) =
\begin{pmatrix}
  \bigoplus_{j=2}^N (0,\phi_j) \\
\phi_1+\bigoplus_{j=2}^N (0,\phi_j) 
\end{pmatrix}
=\begin{pmatrix}
 \bm 0 & K_{N{-}1} \\
 \bm 1 & K_{N{-}1}
\end{pmatrix}\begin{pmatrix}
\phi_1 \\
\bm{\phi}_{2:N}
\end{pmatrix}.
\end{equation}
Hence we obtain the recursion relation in terms of block matrices,
\begin{equation}
K_N=\begin{pmatrix}
 \bm 0 & K_{N{-}1} \\
 \bm 1 & K_{N{-}1}
\end{pmatrix},
\end{equation}
where $\bm 0/\bm 1$ is a column vector of 0/1's of the same number of rows as $K_{N-1}$. This in turn suggests that the rows of $K_N$ are binary-digit vectors of length $N$.

\section{Gate Fidelity estimations}
\subsection{Infidelity upper bound}\label{app:gatefid}
To derive the (average) gate fidelity, we use the formula 
\begin{equation}
 \Fid(\Uideal,\Uqb)= \frac{d+\abs{\tr(\Uideal \Uqb^\dagger )}^2}{d(d+1)},
\end{equation}
where $d=2^N$ is the dimension of the system Hilbert space. 
By construction, $\Uideal$  is diagonal in the computational basis basis,
\begin{equation}\label{eq:Uideal-diag}
\Uideal = \diag(\ee^{-\ii \tau \delta E^{(1)}_1 },\ee^{-\ii \tau \delta E^{(1)}_2 },\cdots,\ee^{-\ii \tau \delta E^{(1)}_d }).
\end{equation}
Hence only the diagonal elements of $\Uqb$ contribute to the gate fidelity. 
The diagonal elements of $\Uqb$ are given by
\begin{equation}\label{eq:Uq-diag}
\matele{n}{\Uqb}{n} = 
\matele{n}{
\ee^{\ii \tau H_0} \ee^{-\ii \tau H}
}{n}
 =\sum_{m} r_{nm}\,  \ee^{-\ii \tau(\wt E_{mn}+\delta E_n)},
\end{equation}
with $r_{nm}=\abs{\braket{n}{\wt m}}^2$,  $\delta E_n \equiv \wt E_n-E_n$ and 
 $\wt E_{mn}\equiv \wt E_m-\wt E_n$. Using \Eqref{eq:Uideal-diag} and \Eqref{eq:Uq-diag}, we acquire the trace product
\begin{equation}\label{eq:trpd}
\tr(\Uideal \Uqb^\dagger) =\sum_{n,m}r_{nm}\,\ee^{-\ii  (\tau \wt E_{mn}+\zeta_n)},
\end{equation}
where we define the excessive phase from 
higher-order energy  corrections
$\zeta_n\equiv \tau(\delta E_n - \delta E_n^{(1)})$.
As $\Uideal$ represents a typical quantum gate, it is expected 
that $\abs{\tau \delta E_n^{(1)}}=O(1)$ and 
$\abs{\zeta_n}\approx\abs{\tau  \delta E_n^{(2)}} \ll 1$.
If only first order perturbation is considered, we have 
$r_{nm}\approx \delta_{nm}$ and $\zeta_n \approx 0$, which produce $d$ for the trace product in \Eqref{eq:trpd} and unit gate fidelity. 
Therefore, the coherent error rate is of second order in the perturbative strength.

To study the contributing factors of the coherent gate error, we now explicitly derive a lower bound of the gate fidelity. 
For easier characterization of  the coherent error strength,
it is better to adopt the gate infidelity $\InF = 1-\Fid$ and 
derive an upper bound for it. Assuming 
$\abs{\tr(\Uideal \Uqb^\dagger)}=d-\epsilon$ for a small deviation $\epsilon\ge 0$, we have
\begin{equation}\label{eq:inf-epsilon}
 \InF(\Uqb, \Uideal) =1-\Fid(\Uqb, \Uideal) 
=  \frac{d^2-(d-\epsilon)^2}{d^2+d} \le \frac{2}{d+1}\epsilon.
\end{equation}
To further upper-bound the gate infidelity, we consider the following series of inequalities
\begin{equation}
 \begin{aligned}
\abs{\tr(\Uideal \Uqb^\dagger)}& \ge 
\abs[\Big]{ \sum_{n,m} r_{nm} \cos\bigl(\tau \wt E_{mn}+ \zeta_n \bigr)}\\
&\ge 
\abs[\Big]{ \sum_{n} r_{nn} \cos(\zeta_n)}
-\abs[\Big]{
\sum_{n} \sum_{m\neq n} r_{nm} \cos\bigl(\tau \wt E_{mn}+\zeta_n\bigr)}\\
&\ge  \sum_{n} r_{nn} \cos(\zeta_n)
- \sum_{n} (1-r_{nn})  \\
& \ge d-2\sum_n(1- r_{nn})-\frac{1}{2}\sum_{n}\zeta_n^2,
\end{aligned}
\end{equation} 
where we have used $\abs{z}\ge\abs{\Re z}$ for the first inequality;
the triangular inequality $\abs{a+b}\ge \abs{a}-\abs{b}$  for the second inequality;
$r_{nm} \ge 0$, together with the normalization relation $\sum_{m} r_{nm} =1$ for the second inequality; 
the trig inequality $\cos x \ge 1-\frac{1}{2}x^2$ together with $r_{nn}\le 1$ for the third inequality. Combined with \Eqref{eq:inf-epsilon}, we obtain the infidelity upper bound 
\begin{equation}
\InF\le \frac{4}{d+1}\sum_n (1-r_{nn})
+ \frac{1}{d+1}\sum_{n}\zeta_n^2,
\end{equation}
appearing in the main text. 

\subsection{The decomposition of gate  errors}\label{app:errdecomp}
Using the second order perturbation theory, one can derive
\begin{align}
  e_\mr{S}^{(2)}&= \sum_n\frac{1}{d}(1-r_{nn}) \simeq \frac{1}{d}\sum_n\sum_{m \neq n}
 \frac{\abs[\big]{ \bra{n} H_\mr{ex} \ket{m} }^2 }{(E_{n}-E_{m})^2},\label{app:eS}\\
 e_\mr{P}^{(2)}
 &\simeq \frac{\tau^2 }{d} \sum_n \Biggl[\sum_{m\neq n} \frac{\abs[\big]{ \bra{n} H_\mr{ex} \ket{m} }^2}{E_n-E_m} \Biggr]^2 .\label{app:eP}
 \end{align}

Let us consider how these error terms can be reduced according to bonds.
Since $H_\mr{ex}$ is a summation of exchange matrices on 2-qubit spaces, the matrix element $\bra{n} H_\mr{ex} \ket{m}\neq 0$ only when the Hamming distance $\mr{D}_{n,m}\le 2$, where the Hamming distance is defined for basis states in the component form, i.e., $\ket{n}=\ket{n_1,n_2,n_3,\cdots,n_N}$ for each component 
$n_i\in\{\uparrow,\downarrow\}$.
For $\mr{D}_{n,m}=2$, supposing $\ket{n}$ and $\ket{m}$ differ in 
the $i$-th and $j$-th component, then 
\begin{align}\label{app:dis2-matele}
\abs[\big]{\bra{n} H_\mr{ex} \ket{m} } &=
\abs[\big]{\bra{n_i,n_j} H_\mr{ex} \ket{\overline{n_i},\overline{n_j}} }=
\begin{cases}
 \frac{1}{2} J_{ij} \abs{s_{ij}}^2 =S_{ij}& \quad \text{if }\ket{n_i}=\ket{n_j}\\
 \frac{1}{2} J_{ij} \abs{t_{ij}}^2 =T_{ij}& \quad \text{if }\ket{n_i}\neq\ket{n_j}
\end{cases},\\[1ex]
E_n-E_m &=E_{n_i,n_j}-E_{\overline{n_i},\overline{n_j}}=
\begin{cases}
  \sig(n_i)(\Ez_i+\Ez_j)& \quad \text{if }\ket{n_i}=\ket{n_j}\\
  \sig(n_i)(\Ez_i-\Ez_j)& \quad \text{if }\ket{n_i} \neq \ket{n_j}
\end{cases}.\label{app:dis2-endiff}
\end{align}
For $\mr{D}_{n,m}=1$, supposing $\ket{n}$ and $\ket{m}$ differ only in 
the $i$-th component, then 
\begin{align}\label{app:dis1-matele}
\abs[\big]{\bra{n} H_\mr{ex} \ket{m}  }^2&=
\abs[\Bigg]{ \sum_{j\neq i} 
\bra{n_i,n_j} H_\mr{ex} \ket{\overline{n_i},n_j}}^2 \notag \\
&=  \sum_{j\neq i}\, \abs[\big]{\bra{n_i,n_j} H_\mr{ex} \ket{\overline{n_i},n_j}}^2
+\sum_{j\neq i} \sum_{k\neq i,j}
{\bra{n_i,n_j} H_\mr{ex} \ket{\overline{n_i},n_j}}^*\,
 \bra{n_i,n_k} H_\mr{ex} \ket{\overline{n_i},n_k}\\[1ex]
E_n-E_m &= \sig(n_i)\, \Ez_i.\label{app:dis1-endiff}
\end{align}

To calculate $e_\mr{S}$ in \Eqref{app:eS}, we can group terms in the summation by the Hamming distance
between $\ket{n}$ and $\ket{m}$.
For the distance-2 sum, we use \Eqref{app:dis2-matele} and \Eqref{app:dis2-endiff} to calculate 
\begin{equation}
\begin{aligned}
 \frac{1}{d}\; \sum_{\mathclap{\substack{n,m\\ \mr{D}_{n,m}=2}}} \frac{\abs[\big]{ \bra{n} H_\mr{ex} \ket{m} }^2 }{(E_{n}-E_{m})^2}
&= \frac{1}{4}\sum_{\langle{i,j}\rangle} 
\sum_{n_i} \sum_{n_j}
\frac{\abs[\big]{ \bra{n_i,n_j} H_\mr{ex} \ket{\overline{n_i},\overline{n_j}} }^2 }{(E_{n_i,n_j}-E_{\overline{n_i},\overline{n_j}})^2}=\sum_{\langle{i,j}\rangle}
\frac{ S_{ij}^2}{2(\Ez_i+\Ez_j)^2} + \frac{T_{ij}^2}{2(\Ez_i-\Ez_j)^2},
\end{aligned}
\end{equation}
where the summation index $\langle{i,j}\rangle$ is for all pairs of directly coupled dots.
For the distance-1 sum, we have according to \Eqref{app:dis1-matele} and \Eqref{app:dis1-endiff},
\begin{align}
 \frac{1}{d}\; &\sum_{\mathclap{\substack{n,m\\ \mr{D}_{n,m}=1}}} \frac{\abs[\big]{ \bra{n} H_\mr{ex} \ket{m} }^2 }{(E_{n}-E_{m})^2}
=\frac{1}{d} \sum_{i=1}^N
\sum_{n_1}\cdots\sum_{n_i}\cdots \sum_{n_N}
\frac{1}{\Ez_i^2}
\abs[\Bigg]{\sum_{j\neq i}
\bra{n_i,n_j} H_\mr{ex} \ket{\overline{n_i},n_j}}^2 \notag \\
&=
\frac{1}{4} \sum_{i} \sum_{j\neq i} \sum_{n_i} \sum_{n_j} 
\frac{\abs[\big]{\bra{n_i,n_j} H_\mr{ex} \ket{\overline{n_i},n_j}}^2}{\Ez_i^2} \notag  \\
&+\frac{1}{8} \sum_{i} \sum_{j\neq i} \sum_{k\neq i,j} \sum_{n_i}\sum_{n_j} 
{\bra{n_i,n_j} H_\mr{ex} \ket{\overline{n_i},n_j}}^*  \underbrace{\sum_{n_k} 
\bra{n_i,n_k} H_\mr{ex} \ket{\overline{n_i},n_k}}_{=0}  \notag  \\
&=\sum_{i} \sum_{j\neq i} \frac{S_{ij}T_{ij}}{\Ez_i^2}
=\sum_{\langle{i,j}\rangle} S_{ij} T_{ij} \bigl(\frac{1}{\Ez_i^2}{+}\frac{1}{\Ez_j^2}\bigr),
\end{align}
where the matrix elements of $H_\mr{ex}$ are calculated according to the definitions \Eqref{eq:ent-state} and \Eqref{eq:Heff}.
Combined with the gate error expression [\Eqref{eq:err2dots}] for a bond, we can conclude that 
\begin{equation}
 e_\mr{S}= \sum_w e^{(w)}_\mr{S}.
\end{equation}

\begin{figure}[htbp]
 \centering
 \includegraphics[width=0.5\linewidth]{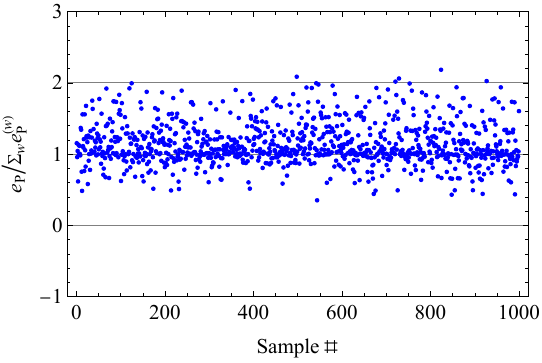}
 \caption{The ratio $e_\mr{P}/\sum_w e^{(w)}_\mr{P}$ for randomly sampled triple-dot systems.}
 \label{fig:eP-ratio}
\end{figure}

The decomposition of phase error $e_\mr{P}$ is trickier than $e_\mr{S}$, as the column and row indices of $H_\mr{ex}$ are not on equal footings in \Eqref{app:eP}. As a result, the full phase error  $e_\mr{P}$ cannot be expressed as the sum of error contributions  $e^{(w)}_\mr{P}$ from all the bonds.  
In \Figref{fig:eP-ratio}, we numerically compute the ratio $e_\mr{P}/\sum_w e^{(w)}_\mr{P}$ for triple-dot systems with randomly sampled quantization energy 
$\Ez_1,\Ez_2,\Ez_3\in [5,10]\ \mu$eV, exchange energy $J_{w}\in[0.01,0.1]\ \mu$eV and complex tunneling coefficients $(s_{w},t_{w})\in[-1-\ii,1+\ii]$ of three possible bonds $w\in\{\langle{1,2}\rangle,\langle{2,3}\rangle,\langle{3,1}\rangle\}$ connecting the dots. 
One can see that for typical parameter range, the full $e_\mr{P}$ is on the same order of magnitude as the sum of bond errors $\sum_w e^{(w)}_\mr{P}$.

We can show that for systems where $\avg{\Ez}\gg \avg{J}$, or if either $s_w=0$ or $t_w=0$ for all the bonds $w$, it holds that $0\le e_\mr{P} \le 2\sum_w e^{(w)}_\mr{P}$. 
If $\avg{\Ez}\gg \avg{J}$,  the dominating terms in the sum \eqref{app:eP} are those with denominator 
$E_n-E_m\propto(\Ez_i-\Ez_j)$, which presents only for distance-2 state pairs $\ket{n},\ket{m}$. Additionally, if $s_w=0$ or $t_w=0$ for all the bonds, the matrix element $\bra{n} H_\mr{ex} \ket{m}=0$  for for all distance-1 state pairs $\ket{n},\ket{m}$. For both cases, we can neglect the distance-1 terms and  approximate 
\begin{equation}
\begin{aligned}
e_\mr{P}&\simeq \frac{\tau^2 }{d} \sum_n 
\Biggl[\sum_{\langle{i,j}\rangle} \frac{\abs[\big]{ \bra{n_i,n_j} H_\mr{ex} \ket{\overline{n_i},\overline{n_j}} }^2}{E_{n_i,n_j}-E_{\overline{n_i},\overline{n_j}}} \Biggr]^2 \le 
\frac{\tau^2 }{d} \sum_n 2\sum_{\langle{i,j}\rangle}
\Biggl[ \frac{\abs[\big]{ \bra{n_i,n_j} H_\mr{ex} \ket{\overline{n_i},\overline{n_j}} }^2}{E_{n_i,n_j}-E_{\overline{n_i},\overline{n_j}}} \Biggr]^2
\\
&=2\, \tau^2 \sum_{\langle{i,j}\rangle} \sum_{n_i}\sum_{n_j}  \frac{\abs[\big]{ \bra{n_i,n_j} H_\mr{ex} \ket{\overline{n_i},\overline{n_j}} }^4}{4(E_{n_i,n_j}-E_{\overline{n_i},\overline{n_j}})^2} \\
&= 2\, \tau^2\Bigl[ \sum_{\langle{i,j}\rangle} \frac{S_{ij}^4}{2(\Ez_i+\Ez_j)^2}  +\frac{T_{ij}^4}{2(\Ez_i- \Ez_j)^2} \Bigr].
\end{aligned}
\end{equation}
On the other hand, from phase error expression for a single bond in \Eqref{eq:err2dots}, we can see that if the cross-terms of $S_wT_w$ can be neglected,  the last line represents the phase error contribution from the bond $\langle i,j \rangle$. Therefore, 
\begin{equation}
e_\mr{P}\lesssim 2\sum_w e_\mr{P}^{(w)}.
\end{equation}
This is in line with the numerical observations in \Figref{fig:eP-ratio}.
\end{document}